\documentclass[sigconf]{acmart}
\usepackage{subfig}
\usepackage{multirow}
\usepackage{pythonhighlight}
\usepackage{threeparttable}
\usepackage{bm}
\usepackage{hyperref}
\AtBeginDocument{%
  \providecommand\BibTeX{{%
    \normalfont B\kern-0.5em{\scshape i\kern-0.25em b}\kern-0.8em\TeX}}}

\setcopyright{acmcopyright}
\copyrightyear{2023}
\acmYear{2023}
\setcopyright{acmlicensed}\acmConference[MM '23]{Proceedings of the 31st
ACM International Conference on Multimedia}{October 29-November 3,
2023}{Ottawa, ON, Canada}
\acmBooktitle{Proceedings of the 31st ACM International Conference on
Multimedia (MM '23), October 29-November 3, 2023, Ottawa, ON, Canada}
\acmPrice{15.00}
\acmDOI{10.1145/3581783.3611694}
\acmISBN{979-8-4007-0108-5/23/10}

\begin{document}

\title{MLIC: Multi-Reference Entropy Model for Learned Image Compression}

\author{Wei Jiang}
\email{wei.jiang1999@outlook.com}
\email{jiangwei@stu.pku.edu.cn}
\orcid{0000-0001-9169-1924}
\affiliation{%
  \institution{Shenzhen Graduate School, Peking University}
  \city{Shenzhen}
  \country{China}
}
\author{Jiayu Yang}
\email{jiayuyang@pku.edu.cn}
\orcid{0000-0001-9729-1294}
\affiliation{%
  \institution{Shenzhen Graduate School, Peking University}
  \institution{Peng Cheng Laboratory}
  \city{Shenzhen}
  \country{China}
}
\author{Yongqi Zhai}
\email{zhaiyongqi@stu.pku.edu.cn}
\orcid{0000-0002-3748-1392}
\affiliation{%
  \institution{Shenzhen Graduate School, Peking University}
  \institution{Peng Cheng Laboratory}
  \city{Shenzhen}
  \country{China}
}
\author{Peirong Ning}
\email{peirongning@stu.pku.edu.cn}
\orcid{0009-0006-6645-5130}
\affiliation{%
  \institution{Shenzhen Graduate School, Peking University}
  \city{Shenzhen}
  \country{China}
}
\author{Feng Gao}
\email{gaof@pku.edu.cn}
\orcid{0009-0006-1843-3180}
\affiliation{%
  \institution{School of Arts, Peking University}
  \city{Beijing}
  \country{China}
}
\author{Ronggang Wang}
\authornote{Corresponding author.}
\orcid{0000-0003-0873-0465}
\email{rgwang@pkusz.edu.cn}
\affiliation{%
  \institution{Shenzhen Graduate School, Peking University}
  \institution{Peng Cheng Laboratory}
  \institution{Migu Culture Technology Co., Ltd}
  \city{Shenzhen}
  \country{China}
}

\renewcommand{\shortauthors}{Wei Jiang et al.}
\begin{abstract}
  Recently, learned image compression has achieved remarkable performance.
    The entropy model, which estimates the distribution of the latent representation,
    plays a crucial role in boosting rate-distortion performance. 
    However, most entropy models only capture correlations in one dimension, 
    while the latent representation contains channel-wise, local spatial, and global spatial correlations. 
    To tackle this issue, we propose the Multi-Reference Entropy Model (MEM) and the advanced version, MEM$^+$.
    These models capture the different types of correlations present in latent representation. Specifically,
    We first divide the latent representation into slices. When decoding the current slice,
    we use previously decoded slices as context and employ the attention map of
    the previously decoded slice to predict global correlations in the current slice.
    To capture local contexts, we introduce two enhanced checkerboard context capturing
    techniques that avoids performance degradation.
    Based on MEM and MEM$^+$, we propose image compression models MLIC and MLIC$^+$.
    Extensive experimental evaluations demonstrate that our MLIC and MLIC+ models achieve state-of-the-art performance,
    reducing BD-rate by $8.05\%$ and $11.39\%$ on the Kodak dataset compared to VTM-17.0 when measured in PSNR.
    Our code is available at \url{https://github.com/JiangWeibeta/MLIC}.
\end{abstract}



\maketitle
\section{Introduction}
Due to the rise of social media, tens of millions of images
are generated and transmitted on the web every second.
Service providers need to find more efficient and effective
image compression methods to save bandwidth.
Although traditional coding methods like JPEG~\cite{pennebaker1992jpeg},
JPEG2000~\cite{DBLP:conf/icmcs/CharrierCL99}, BPG~\cite{bpg}, and VVC~\cite{vtm2019}
have achieved good performance, they rely on manual design for each module
which is independent of each other, making joint optimization impossible.\par
Recently, various learned image compression models
~\cite{DBLP:journals/corr/TodericiOHVMBCS15,balle2017end,
DBLP:conf/icml/RippelB17,DBLP:conf/cvpr/TodericiVJHMSC17,DBLP:conf/iclr/TheisSCH17,
DBLP:conf/cvpr/JohnstonVMCSCHS18,DBLP:conf/cvpr/MentzerATTG18, DBLP:conf/cvpr/LiZGZ018,
DBLP:conf/iclr/LeeCB19, pan2022content, DBLP:conf/aaai/HuY020,DBLP:conf/cvpr/LinYCW20,
DBLP:journals/tip/LiMYZZ20, DBLP:conf/icml/GuoZF021, DBLP:journals/pami/MaLYLW22,
DBLP:journals/corr/abs-2203-10897, koyuncu2022contextformer}
have been proposed, achieved remarkable performance.
Some learned image compression models~\cite{DBLP:conf/cvpr/ChengSTK20,
DBLP:journals/tip/ChenLMSCW21, DBLP:conf/icip/MinnenS20,DBLP:journals/tcsv/WuLZJC22,
DBLP:conf/mm/XieCC21, DBLP:journals/corr/abs-2111-06707} are already comparable to
the advanced traditional method VVC. Most of these models are based on
variational autoencoders~\cite{DBLP:journals/corr/KingmaW13},
following the transform, quantization, entropy coding,
and inverse transform process. Entropy coding plays an important role
in boosting model performance. An entropy model is used to estimate
the entropy of latent representation.
As it involves estimating the entropy of the latent representation using an entropy model.
A powerful and accurate entropy model usually leads to fewer bits.\par
\begin{figure}[htb]
  \centering
  \includegraphics[width=0.95\linewidth]
  {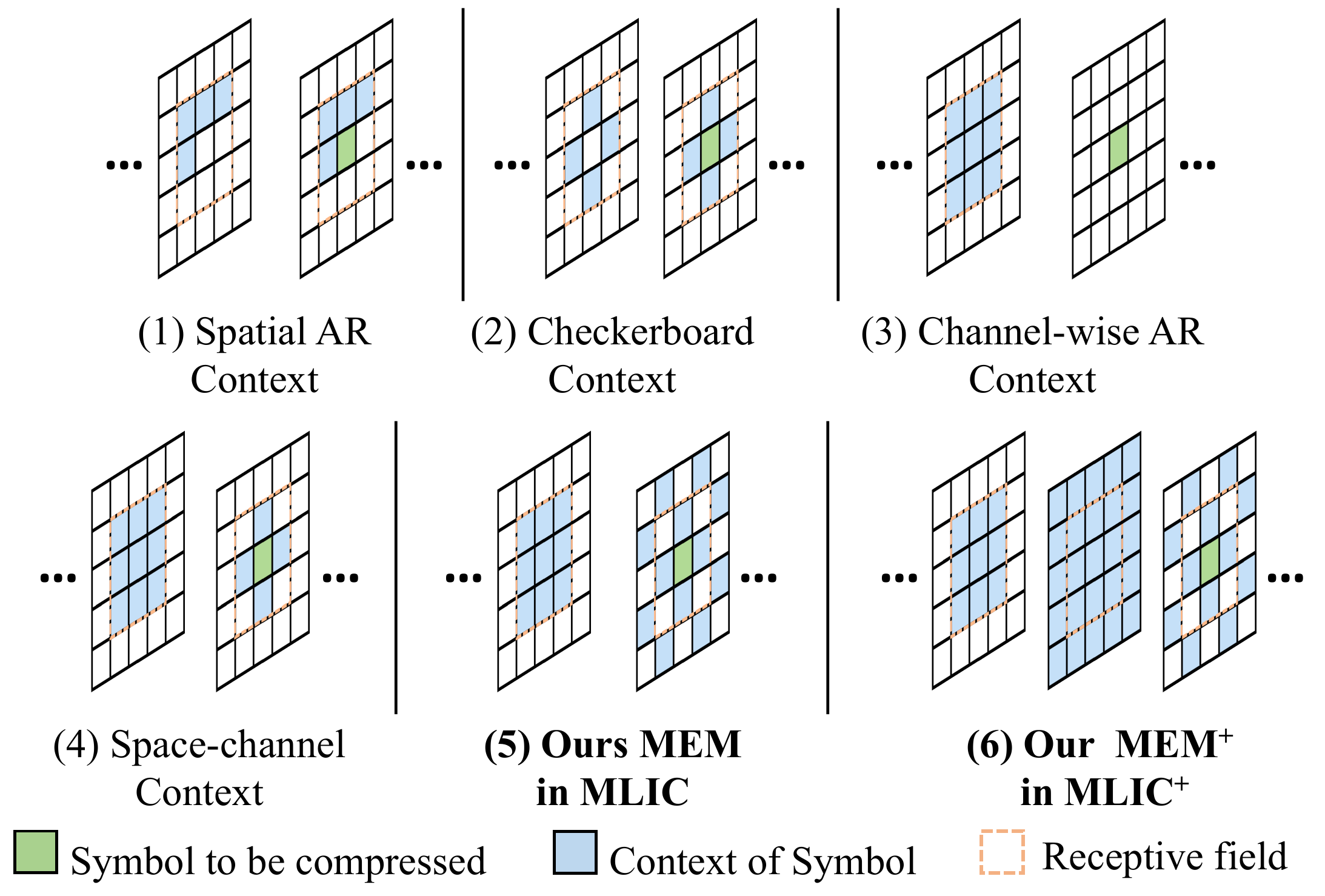}
  \caption{Context Model Comparison. Our proposed Multi-reference Entropy Model MEM and MEM$^+$
  can capture correlations in the local spatial, global spatial, and channel dimensions.
  }
  \label{fig:ctx_compare_new}
\end{figure}
State-of-the-art learned image compression models~\cite{DBLP:conf/iccv/GaoYPHZDL21,
chen2022two, He_2022_CVPR} usually equip the entropy model with a
hyperprior model~\cite{balle2018variational} or a context model~\cite{DBLP:conf/nips/MinnenBT18}
to estimate conditional entropy and use conditional probabilities for coding.
Context models usually model probabilities and correlations in different dimensions,
including local spatial context model,
global spatial context model, and channel-wise context model. However,
most methods capture conditional probabilities in a single dimension,
leading to inaccurate conditional probabilities.
To overcome this limitation,
we introduce Multi-Reference Entropy Models (MEM) and  the enhanced version, MEM+,
which effectively capture local spatial, global spatial, and channel-wise contexts.
Based on MEM and MEM+, we propose MLIC and MLIC+,
which achieve state-of-the-art performance. In our approach, we divide the latent representations into several slices.
When compressing a slice, the previously compressed slices
are treated as its contexts. A channel-wise context module is adopted to
get the channel information from these channel-wise contexts.
We conduct local and global context modeling for every slice separately.
An auto-regressive local context model leads to serial decoding,
while a checkerboard context model~\cite{He_2021_CVPR} can achieve two-pass parallel
decoding which divides latent representations into the anchor and non-anchor parts.
However, the checkerboard context model can result in up to $4\%$ performance degradation.
To address this issue, we propose two different methods:
stacked checkerboard context modeling and overlapped checkerboard window attention.
Some previous methods focused on global context modeling~\cite{DBLP:conf/iclr/QianTSLLSHJ21,
DBLP:journals/tcsv/GuoZFC22} cooperate with serial local spatial context model,
which will cause slower decoding. Assuming similar
spatial correlations in different slices, for the $i$-th slice,
we first compute the attention map of decoded $i-1$-th slice,
which is used to predict the global correlations in $i$-th slice.
We also explore the global correlations in adjacent slices.
Our approach for global spatial context modeling can work well with
checkerboard-like patterns by attention masks. Finally, we fuse channel,
local, global contexts, and side information to compute the
distribution of latent representations.
Our contributions are summarized as follows:
\begin{itemize}
\item We design multi-reference entropy models MEM and MEM$^+$ which
combine local spatial, global spatial and channel contexts, and
hyper-prior side information. Based on MEM and MEM$^+$,
we propose MLIC and MLIC$^+$, which achieve state-of-the-art performance.
We successfully explore the potential of an entropy model.
\item To capture local spatial contexts, we design a stacked
checkerboard context module and checkerboard attention to
address the degradation of checkerboard context modeling while retaining two-pass decoding.
\item We divide latent representation into slices and use the attention map of the
previously decoded slice to predict the global correlations in the current slice.
We also explore global correlations between adjacent slices.
\end{itemize}

\section{Related Works}\label{Sec:related}
\subsection{Learned Lossy Image Compression}
According to rate-distortion optimization, large bit-rate $\mathcal{R}$
leads to lower distortion $\mathcal{D}$.
It is a trade-off. Lagrange multiplier $\lambda$ is used to adjust the weight of
distortion to control the target bit-rate. The optimization target is
\begin{equation}\label{eq:rd}
    \mathcal{L} = \mathcal{R} + \lambda \mathcal{D},
\end{equation}\par
The basic learned image compression framework~\cite{balle2017end} consists
analysis transform $g_a$, quantization $Q$, synthesis transform $g_s$ and
an entropy model to estimate rates. The process can be formulated as:
\begin{equation}
    \boldsymbol{y} = g_a(\boldsymbol{x};\theta), \hat {\boldsymbol{y}} = Q(\boldsymbol{y}), \hat {\boldsymbol{x}} = g_s(\hat {\boldsymbol{y}};\phi),
\end{equation}
where $\boldsymbol x$ is the input image, $g_a$ transform the $\boldsymbol x$ to
compact latent representation $\boldsymbol y$. $\boldsymbol y$
is quantized to $\hat {\boldsymbol{y}}$ for entropy coding.
$\hat {\boldsymbol{x}}$ is the decompressed image.
$\theta$ and $\phi$ are parameters of $g_a$ and $g_s$.
Quantization is non-differentiable, which can be addressed by
adding uniform noise $\mathcal{U}(-0.5, 0.5)$~\cite{balle2017end} or
straight through estimator~\cite{DBLP:conf/iclr/TheisSCH17} during training.
GDN/IGDN~\cite{balle2016gdn} layers are used to improve non-linearity.
In the basic model, a factorized or
a non-adaptive density entropy model is adopted.
The estimated rate of $\hat {\boldsymbol{y}}$ is
$\mathbb{E}[-\log_2p_{\hat {\boldsymbol{y}}}(\hat {\boldsymbol{y}})]$.\par
A hyper-prior model is first introduced in~\cite{balle2018variational},
which extracts side information $\hat {\boldsymbol{z}}$ from $\boldsymbol y$.
Hyper-prior model estimate distribution of $\hat {\boldsymbol{y}}$ from
$\hat {\boldsymbol{z}}$. The rate of $\hat {\boldsymbol{y}}$ is
$\mathbb{E}[-\log_2p_{\hat {\boldsymbol{y}}|\hat {\boldsymbol{z}}}(\hat {\boldsymbol{y}}|\hat {\boldsymbol{z}})]$
and a uni-variate Gaussian distribution model for the hyper-prior is used.
Some works extend it to a mean and scale Gaussian distribution~\cite{DBLP:conf/nips/MinnenBT18},
asymmetric Gaussian distribution~\cite{DBLP:conf/cvpr/CuiWGGFB21} and
Gaussian mixture model~\cite{DBLP:conf/cvpr/ChengSTK20, DBLP:journals/corr/abs-2002-03370}
for more flexible distribution modeling.
\subsection{Context-based Entropy Model}
Many works~\cite{DBLP:conf/nips/MinnenBT18, DBLP:conf/icip/MinnenS20} have been proposed
for more accurate context modeling, including local spatial,
global spatial, and channel-wise context models.\par
Local spatial context models capture correlations between adjacent symbols.
In~\cite{DBLP:conf/nips/MinnenBT18}, a pixel-cnn-like~\cite{van2016pixel} masked convolutional layer
is used to capture local correlations between
$\hat {\boldsymbol{y}}_i$ and symbols $\hat {\boldsymbol{y}}_{<i}$,
which leads to serial decoding.
He \textit{et al.}~\cite{He_2021_CVPR}
divide latent representation $\hat {\boldsymbol{y}}$ into
two parts $\hat {\boldsymbol{y}}_a$ and $\hat {\boldsymbol{y}}_{na}$
and use a checkerboard convolution to extract contexts of
$\hat {\boldsymbol{y}}_{na}$ from $\hat {\boldsymbol{y}}_a$,
achieving two-pass parallel decoding.
\par
Some work aims to model correlations between distant symbols.
In~\cite{DBLP:conf/iclr/QianTSLLSHJ21}, neighbouring left and top symbols
are used as bases for computing the similarity between the target symbol
and its previous symbols. In~\cite{DBLP:journals/tcsv/GuoZFC22},
the latent representation is divided into $2$ parts,
the $L2$ distances of symbols in the first part are used to
predict distant correlations in the second part.
In~\cite{DBLP:journals/corr/abs-2112-04487}, the side information
is divided into global side information and local side information which
leads to extra bits. However, these global context models are incorporated with the serial autoregressive context model,
which further increases decoding latency.\par
Minnen et al.~\cite{DBLP:conf/icip/MinnenS20} model contexts between channels.
$\hat {\boldsymbol{y}}$ is evenly divided to slices. The current slice
$\hat {\boldsymbol{y}}^i$ is conditioned on previously decoded slices
$\hat {\boldsymbol{y}}^{<i}$. An unevenly grouped channel-wise context model is proposed in~\cite{He_2022_CVPR} to
address the uneven distribution of information among slices.\par
\begin{figure}
\centering
\includegraphics[width=0.8\linewidth]
{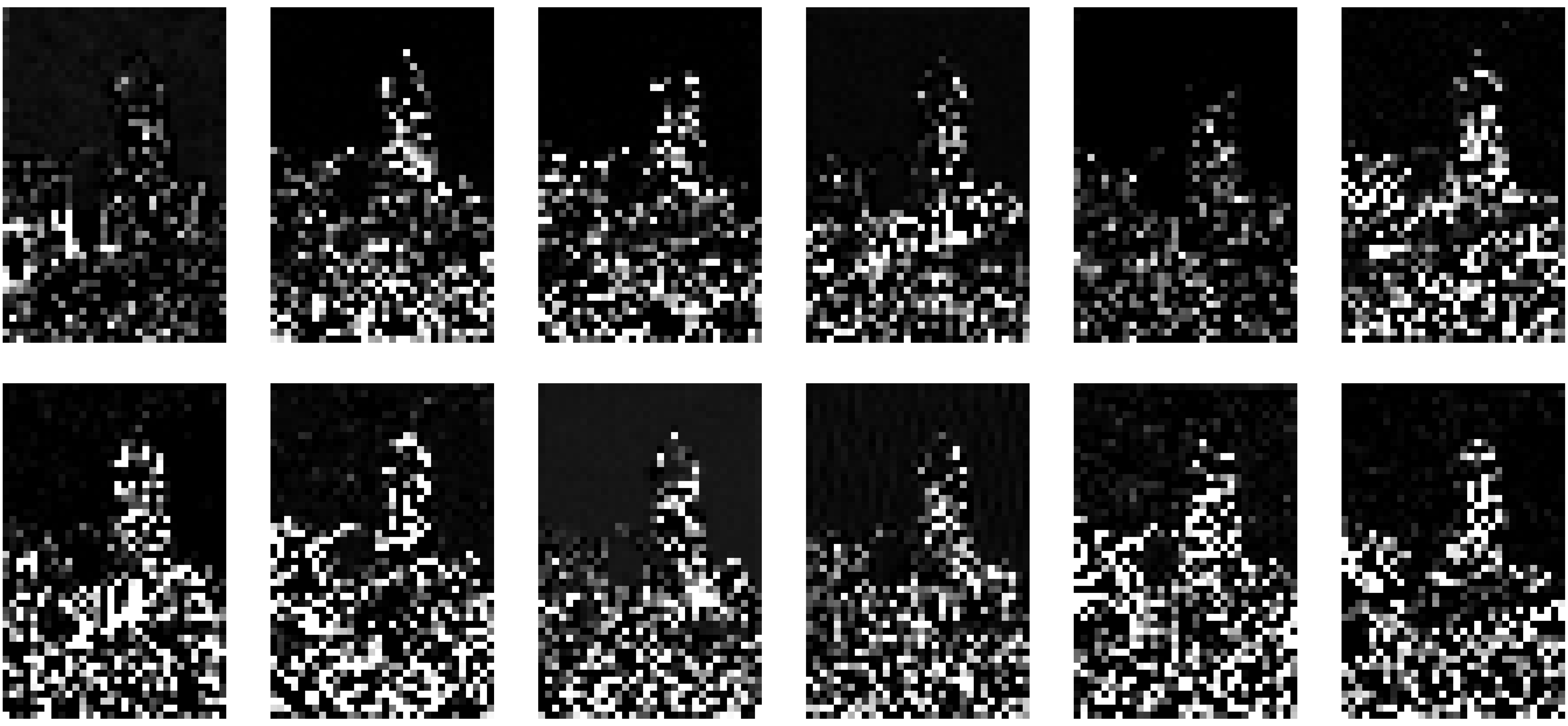}
\caption{Visualization of channels of latent representation of Kodim19
extracted by Cheng'20~\cite{DBLP:conf/cvpr/ChengSTK20}(optimized for MSE, $\lambda=0.0483$).}
\label{cosine}
\end{figure}
While some local and channel-wise context models~\cite{DBLP:journals/corr/abs-2103-02884, He_2022_CVPR}
have demonstrated impressive performance,
effectively capturing local, global, and channel-wise contexts
within a single entropy model remains a challenge.
Addressing these correlations has the potential
to further enhance the performance of the model.
\begin{figure}
\centering
\includegraphics[width=0.55\linewidth]
{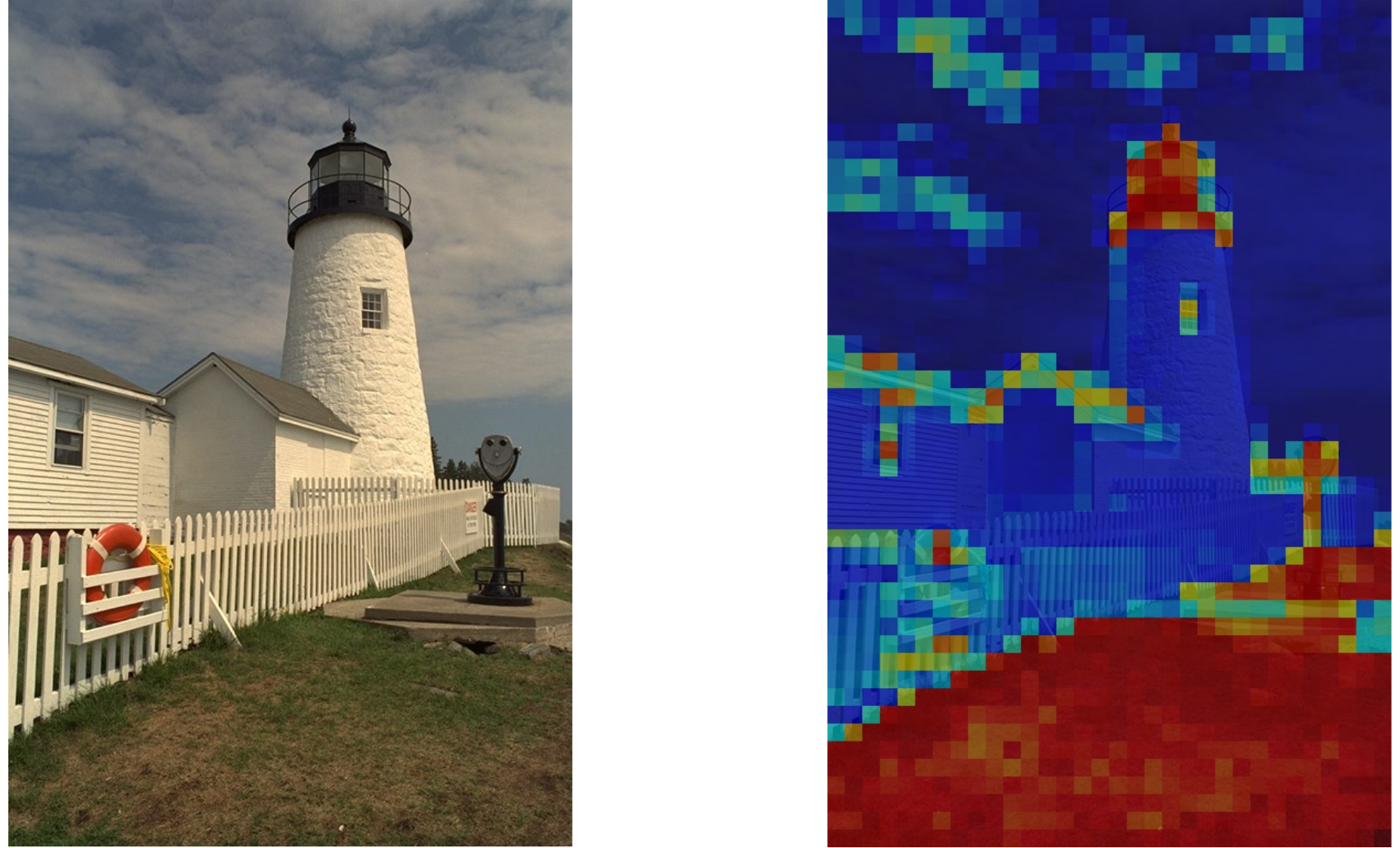}
\caption{Heatmap of spatial cosine similarity of latent representation
of Kodim19 extracted by Cheng'20~\cite{DBLP:conf/cvpr/ChengSTK20} (optimized for MSE, $\lambda=0.0483$).}
\label{heatmap}
\end{figure}
\begin{figure*}
  \centering
  \includegraphics[width=\linewidth]
  {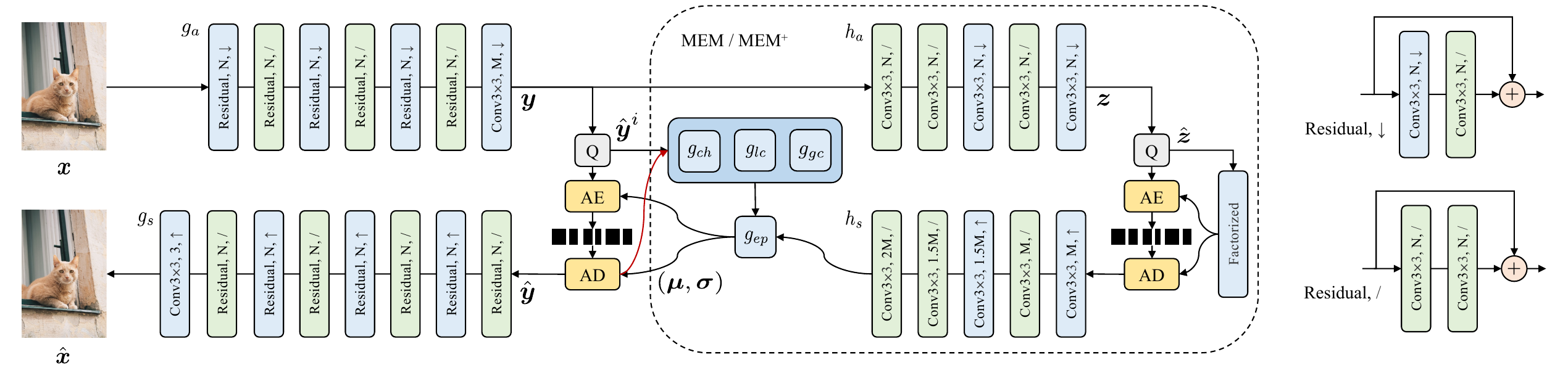}
  \caption{The overall architecture of MLIC and MLIC$^+$.
  $\downarrow$ means down-sampling.
  $\uparrow$ means up-sampling.
  / means stride equals $1$.
  Red line is the dataflow during decoding.
  Please refer to Table~\ref{tab:notation} for the explanations of other notations.}
  \label{fig:arch}
  \end{figure*}
\section{Method}\label{Sec:our method}
\subsection{Motivation}\label{motivation}
According to information theory, conditional entropy is less than or equal to the entropy.
\begin{equation}
    \mathbb{E}[-{\log_2}p_{\hat {\boldsymbol{y}}}(\hat {\boldsymbol{y}})]  \geq \mathbb{E}[-{\log_2}p_{\hat {\boldsymbol{y}}|\boldsymbol{ctx}}(\hat {\boldsymbol{y}}|\boldsymbol{ctx})],
\end{equation}
where $\boldsymbol{ctx}$ is the context of $\hat {\boldsymbol{y}}$.
As long as there are correlations in the latent representation $\hat {\boldsymbol{y}}$,
exploiting these correlations can lead to bit savings.\par
In Figure~\ref{cosine} and Figure~\ref{heatmap},
we first illustrate channel-wise correlations and spatial correlations in
latent representation of Kodim19 extracted by Cheng'20~\cite{DBLP:conf/cvpr/ChengSTK20}.
In Figure~\ref{cosine}, we visualize the feature of several channels.
It's obvious that these features are quite similar.
Capturing such correlations can be challenging for a spatial context model,
since it employs the same mask for all channels when extracting contexts.
This implies that certain correlations may not be fully captured.
In Figure~\ref{heatmap},
symbols with the same color are of high correlation.
A global context model is necessary to capture the correlation
between symbols in the bottom-left corner and those in the bottom-right corner,
where the grass features share similarities.
Although a global context can capture local correlations, a global context can degrade to a local context model,
which make it hard to capture distant correlations because of high similarity among adjacent symbols.
Therefore, we argue that a local context model is necessary.
The latent representation is with redundancy, which means there is potential
to save bits by modeling such correlations. However, previous entropy models
cannot model such correlations both in the spatial and channel domain.
For a spatial context model, the interactions between channels are limited
and for a channel-wise context model, there is no interaction in the current slice,
which inspires us to design multi-reference entropy models.
Our multi-reference entropy model capture correlations in the spatial and channel domain,
which are introduced in the following sections.
\begin{figure}
  \centering
  \includegraphics[width=\linewidth]
  {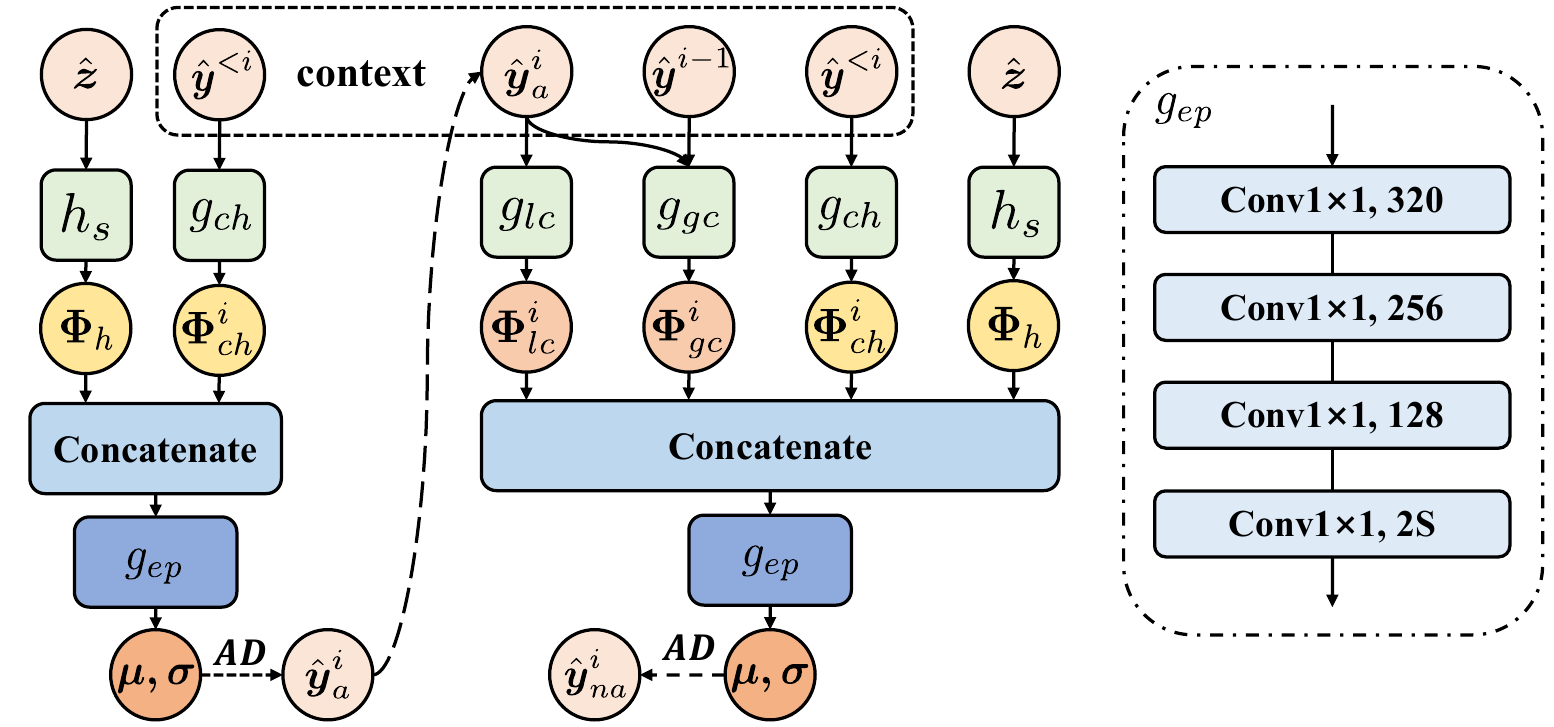}
  \caption{Multi-Reference Entropy Model MEM and MEM$^+$.
  The figure illustrates the process of decoding a slice $\hat {\boldsymbol{y}}^i$.}
  \label{fig:mem}
\end{figure}
\begin{table}
  \footnotesize
  \centering
  \begin{tabular}{c|c}
  \toprule
  Notations                                         & Explanation   \\ \midrule
  $\boldsymbol{x}$, $\hat{\boldsymbol{x}}$          & Input and reconstructed images              \\\midrule
  $\boldsymbol{y}$, $\hat{\boldsymbol{y}}$          & Latent presentation and quantized latent representation               \\\midrule
  $\hat{\boldsymbol{y}}^i, \hat{\boldsymbol{y}}_{a}, \hat{\boldsymbol{y}}_{na}$   & The $i$-th slice of $\hat{\boldsymbol{y}}$, anchor and non-anchor part of $\hat{\boldsymbol{y}}$             \\\midrule
  $\boldsymbol{z}$, $\hat{\boldsymbol{z}}$          & Side information and quantized side information               \\\midrule
  $g_{ep}, \mu,\sigma$                              & Entropy parameter module, Mean and scale of $\hat{\boldsymbol{y}}$           \\\midrule
  $g_a, g_s, h_a, h_s$                              & Analysis and synthesis transform, hyper analysis and synthesis          \\\midrule
  $g_{ch}, g_{lc}, g_{gc}$                          & Channel-wise, local spatial, and global spatial context modules           \\\midrule
  $g_{lc,stk}$                                      & Stacked checkerboard context module     \\\midrule
  $g_{lc,attn}$                                     & Shifted Window-based Checkerboard Attention    \\\midrule
  $g_{gc,intra}, g_{gc,inter}$                      & Intra-slice and Inter-slice global spatial context model     \\\midrule
  $ {\boldsymbol{\Phi}}_{h}, {\boldsymbol{\Phi}_{ch}, {\boldsymbol{\Phi}}_{lc}, {\boldsymbol{\Phi}}_{gc}}$ & hyperprior, channel-wise local spatial, and global spatial context   \\\midrule
  MEM ($^+$)                                        & Multi-reference entropy model ($^+$)             \\\midrule
  $M, N, S, K$                                      & Channel number of $\boldsymbol{y}$, $\boldsymbol{z}$, and $\hat{\boldsymbol{y}}^i$, kernel size            \\\midrule
  Q, AE, AD                                         & Quantization, arithmetic encoding and decoing            \\\midrule
  \end{tabular}
  \caption{Explanations of notations.}
  \label{tab:notation}
\end{table}
\begin{table}
  \footnotesize
  \centering
  \begin{tabular}{lccccccccc}
    \toprule
                & $N$    & $M$    & $S$     & $K$    & Entropy Model       \\ \midrule
      MLIC     & $192$  & $192$  &$32$  & $5$   &   MEM ($g_{lc,stk}, g_{ch}, g_{gc,intra}$)     \\\midrule
      MLIC$^+$ & $192$  & $320$  &$32$  & $5$   &   MEM$^+$ ($g_{lc,attn}, g_{ch}, g_{gc,intra}, g_{gc,inter}$)     \\\midrule
  \end{tabular}
  \caption{Settings of MLIC, MLIC$^+$, MEM, and MEM$^+$.}
  \label{tab:settings}
\end{table}
\subsection{Overview}
\subsubsection{MLIC and MLIC$^+$}
We first give an overview of proposed models MLIC and MLIC$^+$.
The architecture of MLIC and MLIC$^+$ is illustrated in Figure~\ref{fig:arch}.
MLIC and MLIC$^+$ share the same analysis transform $g_a$, synthesis transform $g_s$,
hyper analysis $h_a$ and hyper synthesis $h_s$, which are simplified from
Cheng'20~\cite{DBLP:conf/cvpr/ChengSTK20}.
We remove attention modules to reduce complexity.
The difference between MLIC and MLIC$^+$ is the entropy model.
MLIC is equipped with a Multi-reference entropy model MEM to balance the performance and complexity.
MLIC$^+$ is equipped with a Multi-reference entropy model MEM$^+$
for better rate-distortion performance.
The hyper-parameters and settings of MLIC and MLIC$^+$ are shown in Table~\ref{tab:settings}.
Same to Minnen et al.~\cite{DBLP:conf/icip/MinnenS20},
we adopt \textit{mixed quantization}, meaning adding uniform noise for entropy estimation,
and adopting STE~\cite{DBLP:conf/iclr/TheisSCH17} to make quantization differentiable.
Gaussian mean-scale distribution is adopted for entropy estimation.
\subsubsection{MEM and MEM$^+$}
The Multi-reference entropy models MEM and MEM$^+$ are able to
capture channel-wise, local spatial, and global spatial correlations.
To capture multi-correlations, our MEM and MEM$^+$ contain three parts:
channel-wise context module $g_{ch}$, local spatial context module $g_{lc}$ and
global spatial context module $g_{gc}$.
In the channel-wise context module, the latent representation $\hat{\boldsymbol{y}}$
is divided into slices $\{\hat {\boldsymbol{y}}^0, \hat {\boldsymbol{y}}^1, \cdots\}$.
For the $i$-th slice $\hat {\boldsymbol{y}}^i$, the channel-wise context model captures
the channel-wise context $\boldsymbol{\Phi}_{ch}^i$ from slices $\hat {\boldsymbol{y}}^{<i}$.
To capture local spatial correlations, we adopt the checkerboard pattern, where
the latent representation $\hat{\boldsymbol{y}}$ is divided into
anchor part $\hat{\boldsymbol{y}}_a$ and non-anchor part $\hat{\boldsymbol{y}}_{na}$.
$\hat{\boldsymbol{y}}_a$ is local-context-free.
We capture the local spatial context ${\boldsymbol{\Phi}}_{lc}$ of
$\hat{\boldsymbol{y}}_{na}$ from $\hat{\boldsymbol{y}}_{a}$.
We propose two different approaches:
Stacked Checkerboard Context Module $g_{lc, stk}$ and
Shifted Window-based Checkerboard Attention $g_{lc, attn}$ to capture local spatial contexts.
We divide global spatial contexts $\boldsymbol{\Phi}_{gc}$ into
intra-slice contexts $\boldsymbol{\Phi}_{gc, intra}$, and inter-slice contexts $\boldsymbol{\Phi}_{gc, inter}$.
We propose Intra-Slice Global Context Module $g_{gc, intra}$ and
Inter-Slice Global Context Module $g_{gc, inter}$ to capture such correlations.
We introduce these modules in the following sections.
The structures of MEM and MEM$^+$ are shown in Table~\ref{tab:settings}.
MEM is not equipped with $g_{gc, inter}$ for less complexity.
Following Minnen~\cite{DBLP:conf/icip/MinnenS20},
Latent Residual Prediction modules~\cite{DBLP:conf/icip/MinnenS20} are adopted
to cooperate with channel-wise context module $g_{ch}$.
We illustrate the decompressing process of MLIC and MLIC$^+$ in Figure~\ref{fig:mem}.
We use Equation~\ref{eq:rd} as our loss function
and the estimated rate can be formulated as:
$\mathcal{R} = \sum^L_{i=0} \mathcal{R}_a^i + \mathcal{R}_{na}^i$, where
\begin{equation}
        \mathcal{R}_a^i = \mathbb{E}[-\log_2p_{\hat {\boldsymbol{y}}^i_a | {\boldsymbol{\Phi}}_{ch}^i, {\boldsymbol{\Phi}}_{h}}(\hat {\boldsymbol{y}}^i_a | {\boldsymbol{\Phi}}_{ch}^i, {\boldsymbol{\Phi}}_{h})],
\end{equation}
\begin{equation}
        \mathcal{R}_{na}^i = \mathbb{E}[-\log_2p_{\hat {\boldsymbol{y}}^i_{na} | {\boldsymbol{\Phi}}_{ch}^i, {\boldsymbol{\Phi}}_{h}, {\boldsymbol{\Phi}}_{lc}^i, {\boldsymbol{\Phi}}_{gc}^i}(\hat {\boldsymbol{y}}^i_{na} | {\boldsymbol{\Phi}}_{ch}^i, {\boldsymbol{\Phi}}_{h}, {\boldsymbol{\Phi}}_{lc}^i, {\boldsymbol{\Phi}}_{gc}^i)].
\end{equation}
$\boldsymbol{\Phi}_{h}$ is the hyper-priors extracted by $h_a$ and $h_s$.
\subsection{Channel-wise Context Module}\label{sec:channelctx}
To extract channel-wise contexts, we first evenly divide latent representation
$\hat {\boldsymbol{y}}$ into several slices
$\{\hat {\boldsymbol{y}}^0, \hat {\boldsymbol{y}}^1, \cdots, \hat {\boldsymbol{y}}^L\}$,
$L$ is the number of slices. Slice $\hat {\boldsymbol{y}}^i$ is conditioned on
slices $\hat {\boldsymbol{y}}^{<i}$.
We use a channel context module $g_{ch}$ to squeeze and
extract context information from $\hat {\boldsymbol{y}}^{<i}$ when
encoding and decoding $\hat {\boldsymbol{y}}^i$. $g_{ch}$ consists of
three $3\times 3$ convolutional layers. The channel context becomes
${\boldsymbol{{\boldsymbol{\Phi}}}}_{ch}^i = g_{ch}(\hat {\boldsymbol{y}}^{<i})$.
The channel-wise context module $g_{ch}$ helps select the most relative
channels and extract information beneficial for accurate probability estimation.
The channel number of each slice $S$ is a hyper-parameter.
Following Minnen et al~\cite{DBLP:conf/icip/MinnenS20},
we set $S$ to $32$ in our models.
We adopt latent residual prediction modules~\cite{DBLP:conf/icip/MinnenS20}
to predict quantization error according to
decoded slices and hyper-priors $\boldsymbol{\Phi}_{h}$.
\subsection{Enhanced Checkerboard Context Module}\label{sec:localctx}
The auto-regressive context model $g_{lc, ar}$~\cite{DBLP:conf/nips/MinnenBT18}
leads serial decoding, while the checkerboard context model~\cite{He_2021_CVPR}
makes parallel decoding possible. In the checkerboard context model,
only half of the symbols are conditioned on decoded symbols,
which leads to a slight degradation. We propose two different ways to
solve it from different perspectives.
Note that we capture local spatial contexts for each slice independently.
\begin{figure}
  \centering
  \includegraphics[width=0.9\linewidth]
  {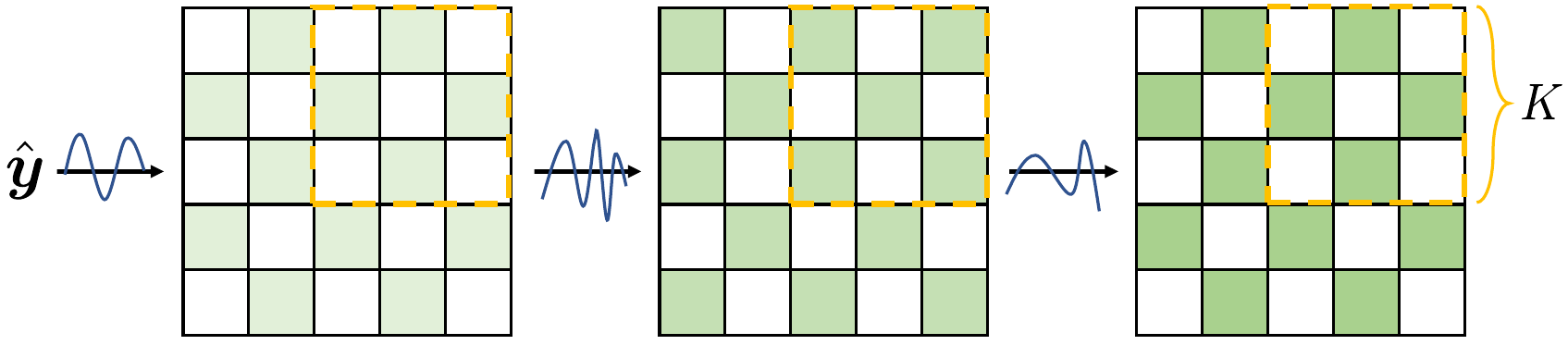}
  \caption{Stacked Checkerboard Context Module $g_{lc, stk}$. The figure illustrate changes of context information after every nonlinear transform. Green squares are context information.}
  \label{stacked_ckbd}
  \end{figure}
\subsubsection{Stacked Checkerboard Context Module}
Depth~\cite{DBLP:conf/cvpr/HeZRS16} and non-linearity are two important factors for
boosting the performance of neural networks.
The deeper and more non-linear the model is, the more expressiveness it has.
In previous work~\cite{DBLP:conf/nips/MinnenBT18, DBLP:journals/tip/LiMYZZ20,
DBLP:conf/cvpr/ChengSTK20, He_2021_CVPR}, local context module is a convolutional layer.
In this module $g_{lc, stk}$, we stack $J$ convolutional layers,
which brings non-linearity and depth. According to the characteristics of the checkerboard pattern,
we then point out that $J$ should be an odd number.
The odd-numbered convolution transfers the information
extracted from the anchor part to the non-anchor part and the even-numbered
convolution transfers the information extracted from the non-anchor part to the anchor part.
The transfer process of context information is shown in Figure~\ref{stacked_ckbd}
when $J$ is $3$. We set $J = 3$ to balance model performance and parameters
and set kernel size $K=5$.
\begin{figure}
  \centering
  \includegraphics[width=\linewidth]
  {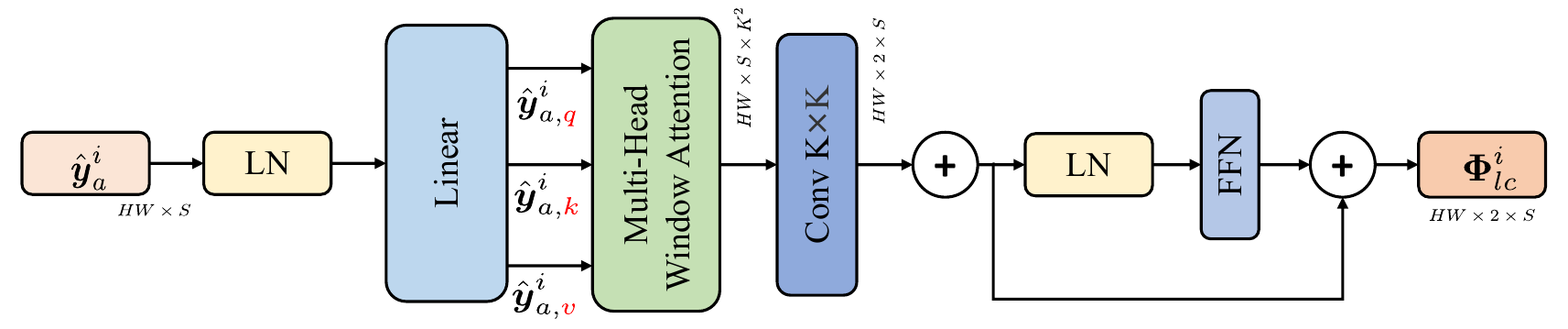}
  \caption{Checkerboard Attention Context Module $g_{lc, attn}$.}
  \label{fig:ckbd_attn_arch}
\end{figure}
\begin{figure}
\centering
\includegraphics[width=0.9\linewidth]
{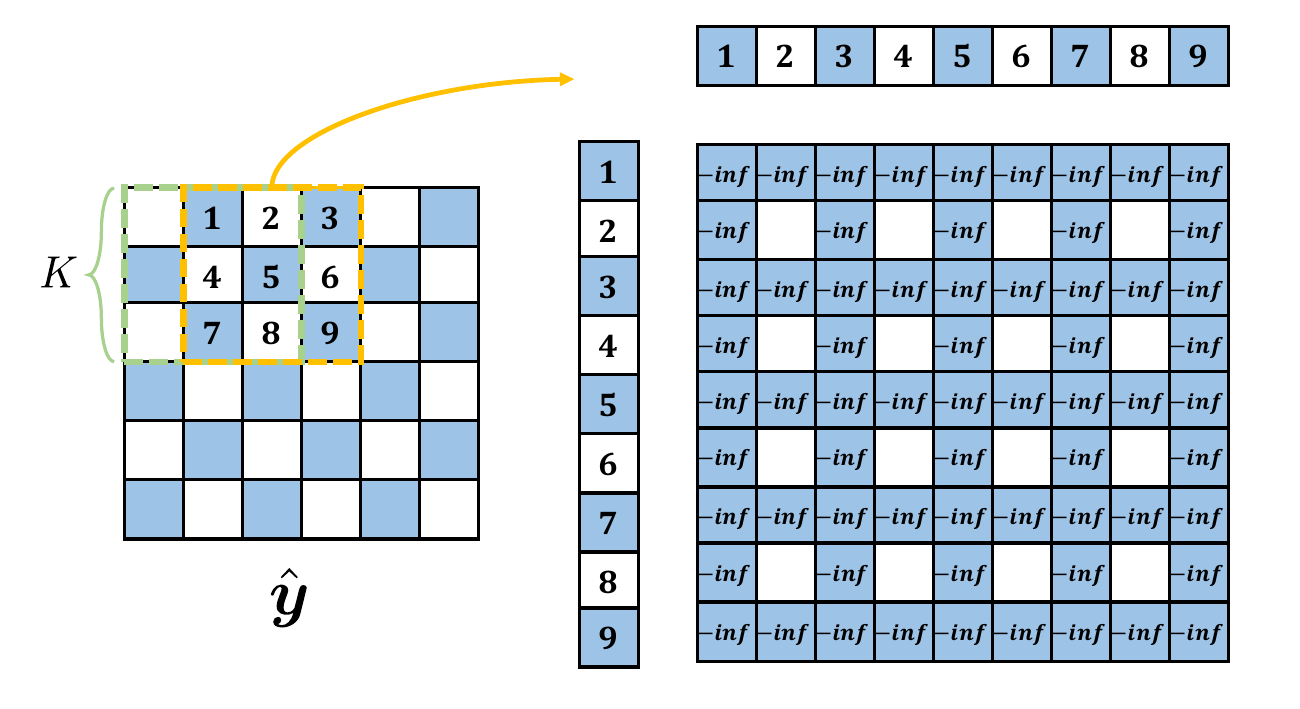}
\caption{Mask of Shifted Window-based Checkerboard Attention $g_{lc, attn}$.
Blue squares are non-anchor part $\hat {\boldsymbol{y}}_{na}$,
white squares are anchor part $\hat {\boldsymbol{y}}_a$.
Green square and yellow square are two windows.}
\label{ckbd_attn}
\end{figure}
\subsubsection{Shifted-Window-based Checkerboard Attention}
One drawback of the CNN-based local context module is the fixed weights,
which makes them impossible to capture content-adaptive contexts.
We contend that context-adaptation is beneficial due to the wide range of image diversity.
In transformers~\cite{DBLP:conf/nips/VaswaniSPUJGKP17, DBLP:conf/iccv/LiuL00W0LG21},
the attention weight is generated dynamically according to the input,
which inspires us to design a transformer-based content-adaptive local context module.
The local receptive field is like a window,
we capture the local spatial contexts by dividing the feature map into overlapped windows.
We propose the checkerboard attention context module $g_{lc, attn}$.
We take the $i$-th slice as example.
Assumimg the resolution of the latent representation $\hat {\boldsymbol{y}}^i$ is $H \times W$,
we divide $\hat {\boldsymbol{y}}^i$
into $H \times W$ overlapped windows and the window size is $K\times K$.
To extract local correlations, we first compute the attention map of each window.
Same as the convolutional checkerboard context model, interactions between $\boldsymbol{y}_{a}^i$ and $\boldsymbol{y}_{na}^i$
and interactions in $\boldsymbol{y}_{na}^i$ are not allowed. An example of the attention mask is illustrated
in Figure~\ref{ckbd_attn}. Such attention does not change the resolution of each window.
We use a $K\times K$ convolutional layer to fusion local context information and
make the size of the local context same as that of $\boldsymbol{y}^i$ before feeding
it to an FFN~\cite{DBLP:conf/nips/VaswaniSPUJGKP17}. The process be formulated as:
\begin{equation}
    {{\hat {\boldsymbol{y}}}^i}_{attn} = \textrm{softmax}\left(\frac{{{\hat {\boldsymbol{y}}}^i}_{a,q} \times ({{\hat {\boldsymbol{y}}}^i}_{a,k})^\top}{\sqrt{S}} + mask\right) \times {{\hat {\boldsymbol{y}}}^i}_{a,v},
\end{equation}
\begin{equation}
    {{\hat {\boldsymbol{y}}}^i}_{conv} = \textrm{conv}_{K\times K}({{\hat {\boldsymbol{y}}}^i}_{attn}),
\end{equation}
\begin{equation}
    {\boldsymbol{\Phi}}^i_{lc} = \textrm{FFN}({{\hat {\boldsymbol{y}}}^i}_{conv}) + {{\hat {\boldsymbol{y}}}^i}_{conv},
\end{equation}
where $\hat {\boldsymbol{y}}^i_{a,q}, \hat {\boldsymbol{y}}^i_{a,k}, \hat {\boldsymbol{y}}^i_{a,v} = \textrm{Embed}(\hat {\boldsymbol{y}}^i_a)$,
$\hat {\boldsymbol{y}}^i_a$ is anchor part of $i$-th slice, $mask$ the attention mask,
$S$ is the channel number of each slice.\par
Note that our overlapped window-partition is with linear complexity.
The complexity of $g_{lc, attn}$ is $\Omega(2K^4HWS + 4HWS^2)$,
where $S$ is the channel number of a slice.
\subsection{Global Spatial Context Module}\label{sec:globalctx}
\begin{figure}
  \centering
  \includegraphics[width=0.85\linewidth]
  {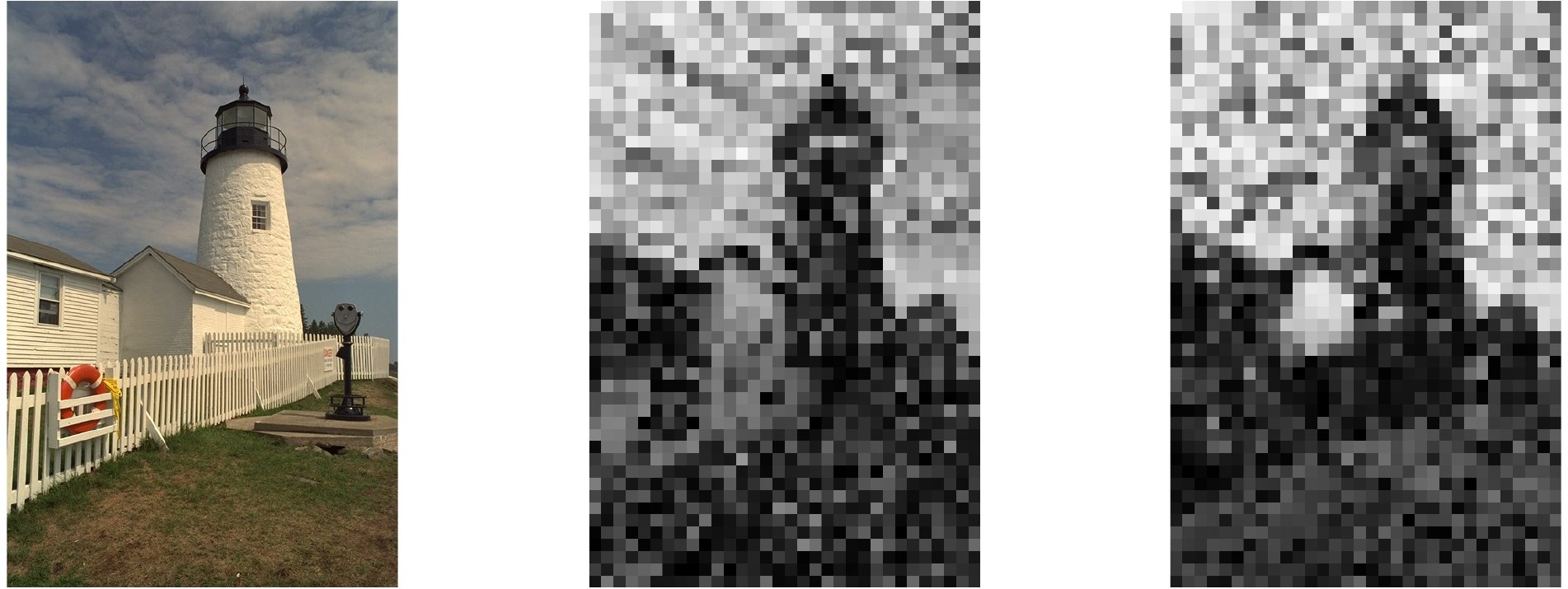}
  \caption{Cosine similarity in the spatial domain of different slices of latent representation of Kodim19 extracted by Cheng'20~\cite{DBLP:conf/cvpr/ChengSTK20} (optimized for MSE).}
  \label{intra_cosine}
\end{figure}
We capture global-spatial contexts for each slice independently.
We explore global correlations between non-anchor part $\hat {\boldsymbol{y}}_{na}^i$
and anchor part $\hat {\boldsymbol{y}}_a^i$ in a slice $\hat {\boldsymbol{y}}^i$ and
between non-anchor part $\hat {\boldsymbol{y}}_{na}^i$ and slice
$\hat {\boldsymbol{y}}^{i-1}$.
\begin{figure}
  \centering
  \includegraphics[width=\linewidth]
  {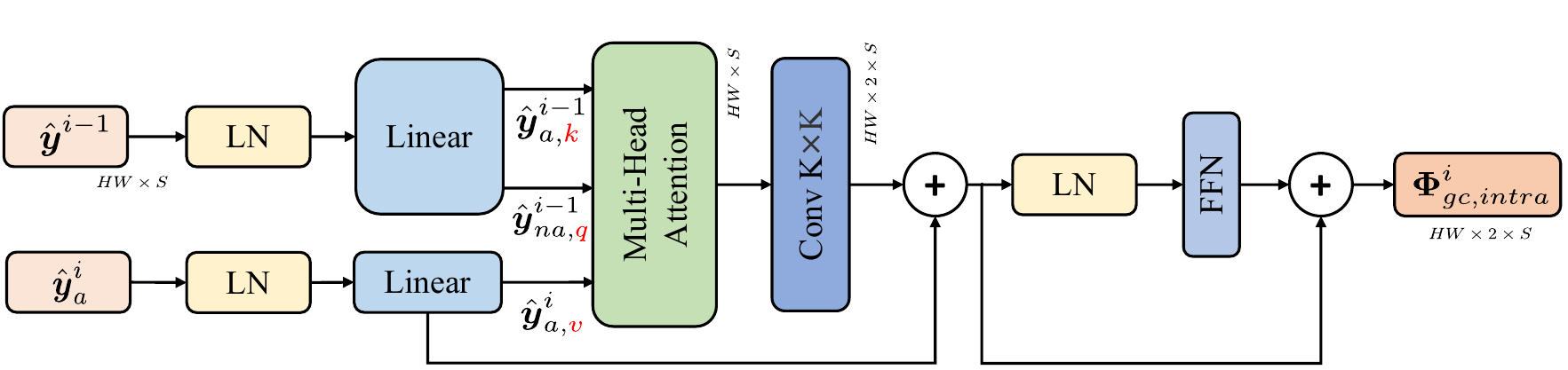}
  \caption{Intra-Slice Global Context Model $g_{gc, intra}$. $S$ is the channel number of a slice.}
  \label{intra_arch}
  \end{figure}
  \begin{figure}
  \centering
  \includegraphics[width=\linewidth]
  {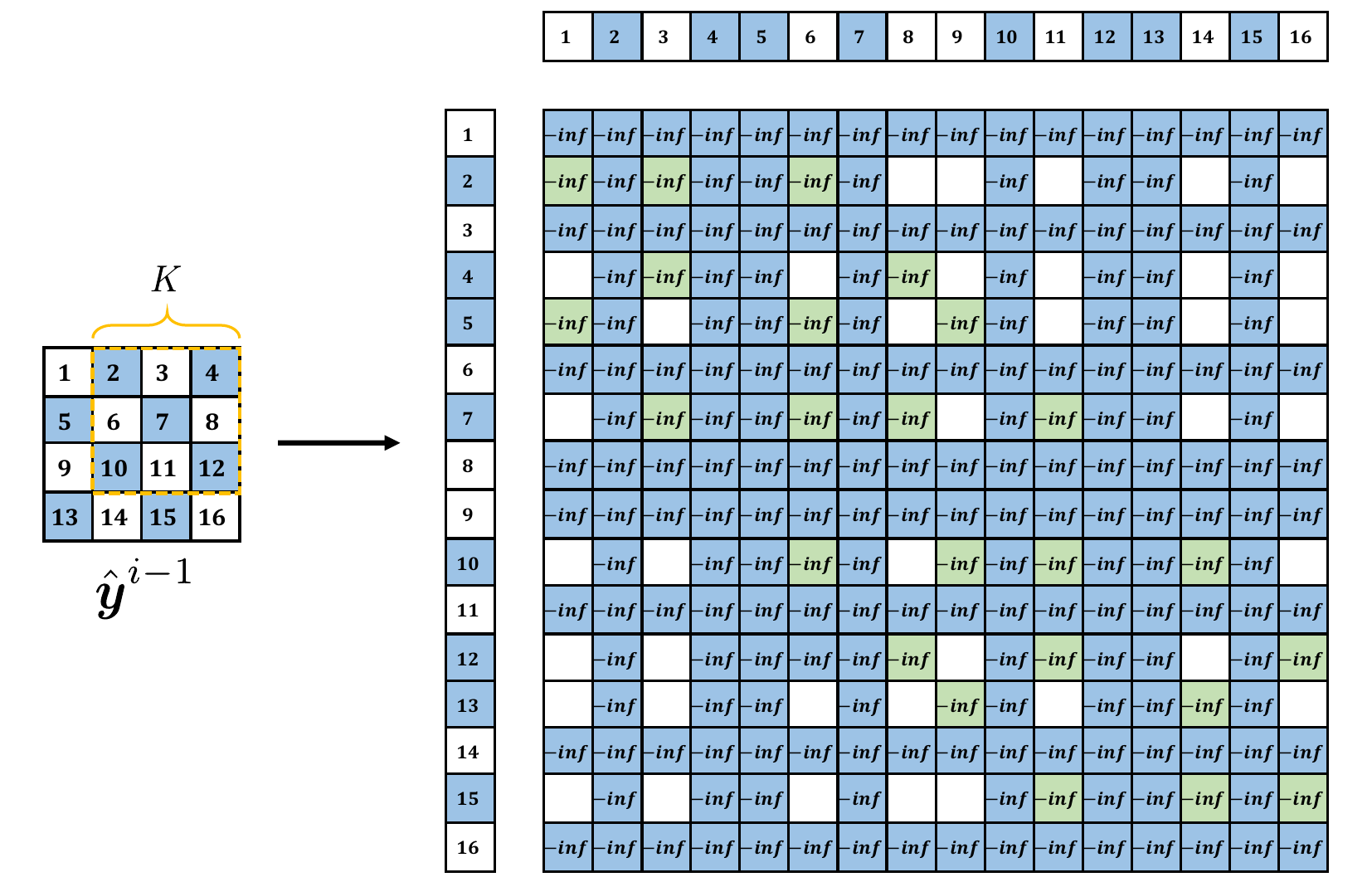}
  \caption{Mask of Intra-Slice Global Context Model $g_{gc, intra}$.
  Blue squares belong to non-anchor part $\hat {\boldsymbol{y}}^{i-1}_{na}$ and
  white squares belong to anchor part $\hat {\boldsymbol{y}}^{i-1}_a$ of
  slice $\hat {\boldsymbol{y}}^{i-1}$. The green squares are masked to avoid
  interactions in the local receptive field.
  The orange dotted box is the receptive field of the local context model.}
  \label{intra_attn}
\end{figure}
\begin{figure*}
  \centering
  \subfloat{
    \includegraphics[scale=0.47]{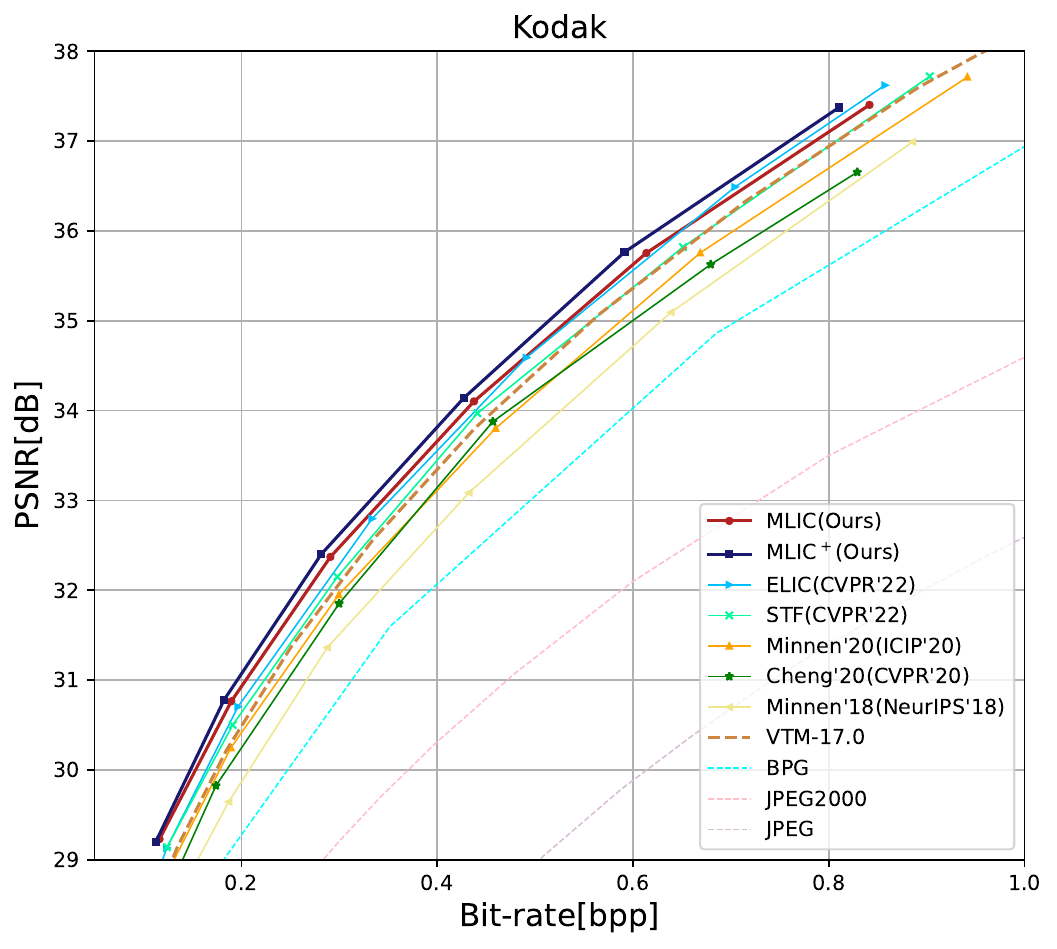}}
  \subfloat{
    \includegraphics[scale=0.47]{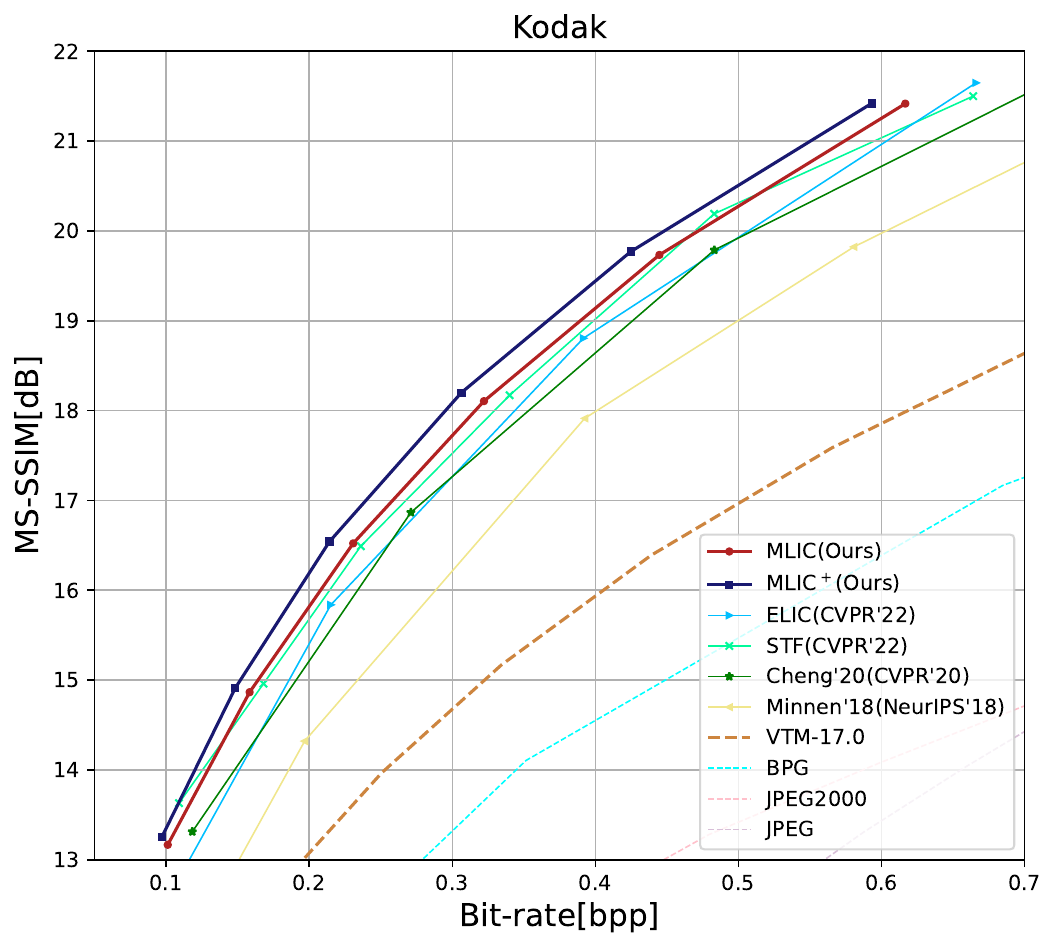}}
  \caption{PSNR-Bit-rate curve (opt.MSE) and MS-SSIM-Bit-rate curve (opt.MS-SSIM) on Kodak dataset.}
  \label{rd}
\end{figure*}
\subsubsection{Intra-Slice Global Context Module}
Due to codec consistency, it is impossible to know the global correlations
between current symbols and other symbols during decoding.
One solution is writing global correlations into bit-stream,
which causes extra bits.
In latent representation $\hat{\boldsymbol{y}}$, each channel contains different information
but each channel can be treated as a thumbnail.
We point out that channels share similar global similarities.
We illustrate the cosine similarity of two slices
of Cheng'20~\cite{DBLP:conf/cvpr/ChengSTK20} in Figure~\ref{intra_cosine}.
Their global correlations are similar despite differences in magnitude.
When decoding the current slice $\hat {\boldsymbol{y}}^i$, decoded slice
$\hat {\boldsymbol{y}}^{i-1}$ helps estimate the global correlations
in slice $\hat {\boldsymbol{y}}^{i}$.
One problem is how to eatimate the global correlations.
Cosine similarity may be helpful, however, it is fixed and may not be accurate for feature.
We point out attention map is a good choice.
The embedding layer is learnable, which make it flexible
to adjust the method for global correlations estimation by changing queries, keys, and values.
We take the $i-1$-th slice and the $i$-th as example.
When compressing or decompressing $\hat {\boldsymbol{y}}^i$,
we first compute the correlations between anchor part $\hat {\boldsymbol{y}}^{i-1}_a$
and non-anchor part $\hat {\boldsymbol{y}}^{i-1}_{na}$ of slice $\hat {\boldsymbol{y}}^{i-1}$,
because the checkerboard local context model makes anchor visible when decoding non-anchor part.
We multiply the anchor part of current slice $\hat {\boldsymbol{y}}^{i}_{a}$ with
an attention map.
Due to the local correlations, adjacent symbols have similar global correlations.
We use a $K\times K$ convolutional layer to refine the attention
map by aggregating global similarities of adjacent symbols.
The process of this Intra-Slice Global Context $g_{gc, inter}$ can be formulated as:
\begin{equation}
    \hat {\boldsymbol{y}}^{i}_{attn} = \textrm{softmax}\left(\frac{{{\hat {\boldsymbol{y}}}^{i-1}}_{na,q} \times ({{\hat {\boldsymbol{y}}}^{i-1}}_{a,k})^\top}{\sqrt{S}} + mask\right) \times \hat {\boldsymbol{y}}^{i}_{a,v},
\end{equation}
\begin{equation}
    \hat {\boldsymbol{y}}^{i}_{conv} = \textrm{conv}_{K\times K}(\hat {\boldsymbol{y}}^{i}_{attn}) + \hat {\boldsymbol{y}}^{i}_{a,v},
\end{equation}
\begin{equation}
    {\boldsymbol{\Phi}}^i_{gc,intra} = \textrm{FFN}(\hat {\boldsymbol{y}}^{i}_{conv}) + \hat {\boldsymbol{y}}^{i}_{conv},
\end{equation}
where $\hat {\boldsymbol{y}}^{i-1}_{na,q}, \hat {\boldsymbol{y}}^{i-1}_{a,k} = \textrm{Embed}(\hat {\boldsymbol{y}}^{i-1})$, $\hat {\boldsymbol{y}}^{i}_{a,v} = \textrm{Embed}(\hat {\boldsymbol{y}}^i_{a})$.
Note that interactions within $\hat {\boldsymbol{y}}^{i-1}_{na}$ and $\hat {\boldsymbol{y}}^{i-1}_a$
are masked. Local receptive fields are also masked.
We point out that if we don't adopt the mask, the intra-slice global context model
can degrade to local context model becuase high similarities between adjacent symbols.
Local correlations may dominate the attention map.
The $mask$ is illustrated in Figure~\ref{intra_attn}.
\subsubsection{Inter-Slice Global Context Module}
Because of the global correlations between slices,
we extend the intra-slice global context to the inter-slice global context.
We explore the correlations between $\hat {\boldsymbol{y}}^{i}_{na}$ and $\hat {\boldsymbol{y}}^{i-1}$
by using $\hat {\boldsymbol{y}}^{i}_a$ as an approximation of $\hat {\boldsymbol{y}}^{i}_{na}$.
We only explore the global correlations between adjacent slices to control complexity.
In inter-slice global context module, we also adopt
the learnable attention map and a convolutional layer for refinement.
A mask is adopted to aviod interactions between $\hat {\boldsymbol{y}}^{i}_{na}$ and $\hat {\boldsymbol{y}}^{i-1}$.
The process of this Inter-Slice Global Context Model $g_{gc, inter}$ can be formulated as:
\begin{equation}
  \hat {\boldsymbol{y}}^{i}_{attn} = \textrm{softmax}\left(\frac{{{\hat {\boldsymbol{y}}}^{i}}_{a,q} \times ({{\hat {\boldsymbol{y}}}^{i-1}}_{k})^\top}{\sqrt{S}} + mask\right) \times \hat {\boldsymbol{y}}^{i-1}_{v},
\end{equation}
\begin{equation}
  \hat {\boldsymbol{y}}^{i}_{conv} = \textrm{conv}_{K\times K}(\hat {\boldsymbol{y}}^{i}_{attn}) + \hat {\boldsymbol{y}}^{i-1}_{v},
\end{equation}
\begin{equation}
  {\boldsymbol{\Phi}}^i_{gc,inter} = \textrm{FFN}(\hat {\boldsymbol{y}}^{i}_{conv}) + \hat {\boldsymbol{y}}^{i}_{conv},
\end{equation}
where $\hat {\boldsymbol{y}}^{i-1}_{k}, \hat {\boldsymbol{y}}^{i-1}_{v} = \textrm{Embed}(\hat {\boldsymbol{y}}^{i-1})$, $\hat {\boldsymbol{y}}^{i}_{a,q} = \textrm{Embed}(\hat {\boldsymbol{y}}^i_{a})$.
\section{Experiments}\label{Sec:exp}
\subsection{Settings}
We select $2\times 10^5$ images from COCO2017~\cite{lin2014microsoft},
DIV2K~\cite{Agustsson_2017_CVPR_Workshops}, ImageNet~\cite{ILSVRC15}
with a resolution larger than $480\times 480$ as our training set.
We train our model on a single Tesla V100 GPU with various configurations
of the Lagrange multiplier $\lambda$ for different quality presets.
We use MSE and MS-SSIM as distortion metrics.
Following the settings of CompressAI~\cite{DBLP:journals/corr/abs-2011-03029},
we set $\lambda \in \{18, 35, 67, 130, 250, 483\} \times 10^{-4}$
for MSE and $\lambda \in \{2.40, 4.58, 8.73, 16.64, 31.73, 60.50\}$
for MS-SSIM~\cite{wang2003multiscale}.
We train each model with an Adam optimizer with $\beta_1=0.9, \beta_2=0.999$
and the batch size is 8. We train each model for 2M steps.
The learning rate starts at $10^{-4}$ and drops to $3\times 10^{-5}$ at 1.5M steps,
drops to $10^{-5}$ at 1.8M steps, and drops to $3 \times 10^{-6}$ at 1.9M steps,
drops to $10^{-6}$ at 1.95M steps.
During training, we random crop images to $256\times 256$ patches during the
first $1.2$M steps, and crop images to $448\times 448$ patches
during the rest steps due to the sparsity of intra-slice and
inter-slice attention masks shown in Figure~\ref{intra_attn}.
Large patches are beneficial for models to learn global references.
\begin{table*}[t]
  \footnotesize
  \centering
  \begin{tabular}{@{}cccccccccccccc@{}}
  \toprule
  \multicolumn{1}{c|}{\multirow{2}{*}{Methods}}                            & \multicolumn{2}{c}{Kodak~\cite{kodak}}       & \multicolumn{2}{c}{Tecnick~\cite{asuni2014testimages}}   & \multicolumn{2}{c}{CLIC Pro Val~\cite{clic2020dataset}} & \multicolumn{2}{c}{CLIC'21 Test~\cite{clic2021dataset}} & \multicolumn{2}{c}{CLIC'22 Test~\cite{clic2022dataset}}  & \multicolumn{2}{c}{JPEGAI Test~\cite{jpegai}}   \\
  \multicolumn{1}{c|}{}                                                     & \multicolumn{1}{c}{PSNR} & \multicolumn{1}{c}{MS-SSIM} & \multicolumn{1}{c}{PSNR}& \multicolumn{1}{c}{MS-SSIM} & \multicolumn{1}{c}{PSNR}& \multicolumn{1}{c}{MS-SSIM}& \multicolumn{1}{c}{PSNR}& \multicolumn{1}{c}{MS-SSIM} & \multicolumn{1}{c}{PSNR}& \multicolumn{1}{c}{MS-SSIM} & \multicolumn{1}{c}{PSNR}& \multicolumn{1}{c}{MS-SSIM}\\ \midrule
  \multicolumn{1}{c|}{VTM-17.0~\cite{vtm2019}}                                        & $0.00$       & $0.00$         & $0.00$         & $0.00$  & $0.00$    & $0.00$    & $0.00$    & $0.00$  & $0.00$    &$0.00$    &$0.00$    &$0.00$ \\\midrule
  \multicolumn{1}{c|}{Cheng'20 (CVPR'20)~\cite{DBLP:conf/cvpr/ChengSTK20}}            & $+5.58$      & $-44.21$       & $7.57$       & $-39.61$ & $+11.71$ & $-41.29$ & $+9.40$ & $-37.22$ & $+13.29$ & $-33.40$ & $+11.95$ & $-40.03$ \\\midrule
  \multicolumn{1}{c|}{Minnen'20 (ICIP'20)~\cite{DBLP:conf/icip/MinnenS20}}            & $+3.23$      & $--$            & $-0.88$        & $--$     & $--$  & $--$  & $--$  & $--$  & $--$  & $--$  & $--$  & $--$ \\\midrule
  \multicolumn{1}{c|}{Qian'21 (ICLR'21)~\cite{DBLP:conf/iclr/QianTSLLSHJ21}}          & $+10.05$      & $-39.53$       & $--$       & $--$  & $--$       & $--$ & $--$       & $--$ & $--$       & $--$  & $--$  & $--$  \\\midrule
  \multicolumn{1}{c|}{Xie'21 (MM'21)~\cite{DBLP:conf/mm/XieCC21}}                     & $+1.55$      & $-43.39$       & $-0.80$       & $--$ &$+3.21$ &  $--$& $+0.99$  & $--$ & $+2.13$ &  $--$ & $+2.35$ & $--$\\\midrule
  \multicolumn{1}{c|}{Entroformer (ICLR'22)~\cite{DBLP:journals/corr/abs-2202-05492}} & $+4.73$      & $-42.64$       & $+2.31$              &$--$      &$--$  &$--$  &$--$  &$--$  &$--$   &$--$  &$--$  &$--$  \\\midrule
  \multicolumn{1}{c|}{SwinT-Charm (ICLR'22)~\cite{zhu2021transformer}}                & $-1.73$      & $-42.64$       & $+6.50$     & $--$ & $--$  & $--$ &$+2.56$   &$--$ &$--$ &$--$ & $+3.16$ & $--$  \\\midrule
  \multicolumn{1}{c|}{WACNN (CVPR'22)~\cite{DBLP:journals/corr/abs-2203-08450}}       & $-2.95$      & $-47.71$       & $--$        & $--$  & $+0.04$ & $-44.38$ & $--$ & $--$ & $--$ & $--$  & $--$ & $--$\\\midrule
  \multicolumn{1}{c|}{STF (CVPR'22)~\cite{DBLP:journals/corr/abs-2203-08450}}         & $-2.48$      & $-47.72$       & $-2.75$        & $--$  & $+0.42$ & $-44.82$ & $-0.16$ &$--$ &$+0.08$ & $--$  &$+1.54$ & $--$\\\midrule
  \multicolumn{1}{c|}{ELIC (CVPR'22)~\cite{He_2022_CVPR}}                             & $-5.95$      & $-44.60$       & $--$        & $--$ & $--$ & $--$ & $-7.52$ & $--$ & $--$ & $--$ & $--$ & $--$\\\midrule
  \multicolumn{1}{c|}{NeuralSyntax (CVPR'22)~\cite{Wang_2022_CVPR}}                   & $+8.97$      & $-39.56$       & $--$        & $--$  & $+5.64$ &$-38.92$ & $--$ & $--$ & $--$ & $--$ & $--$ & $--$\\\midrule
  \multicolumn{1}{c|}{Informer (CVPR'22)~\cite{DBLP:journals/corr/abs-2112-04487}}    & $+10.01$     & $-39.25$       & $+9.72$        & $--$ & $--$ & $--$ & $--$ & $--$ & $--$ & $--$ & $--$ & $--$\\\midrule
  \multicolumn{1}{c|}{McQuic (CVPR'22)~\cite{DBLP:journals/corr/abs-2203-10897}}      & $-1.57$      & $-47.94$       & $--$        & $--$ & $+6.82$ & $-40.17$ & $--$ & $--$ & $--$ & $--$ & $--$ & $--$\\\midrule
  \multicolumn{1}{c|}{Contextformer (ECCV'22)~\cite{koyuncu2022contextformer}}        & $-5.77$      & $-46.12$       & $-9.05$        & $-42.29$ & $--$ & $--$ & $--$ & $--$ & $--$ & $--$ & $--$ & $--$ \\\midrule
  \multicolumn{1}{c|}{Pan'22 (ECCV'22)~\cite{pan2022content}}                         & $+7.56$      & $-36.20$       & $+3.97$        & $--$ & $--$ & $--$ & $--$ & $--$ & $--$ & $--$ & $--$ & $--$  \\\midrule
  \multicolumn{1}{c|}{MLIC (Ours)}                                                    & \textcolor{blue}{\bm{$-8.05$}}      &\textcolor{blue}{\bm{$-49.13$}}      &\textcolor{blue}{\bm{$-12.73$}}  &\textcolor{blue}{\bm{$-47.26$}} & \textcolor{blue}{\bm{$-8.79$}} &\textcolor{blue}{\bm{$-45.79$}} &\textcolor{blue}{\bm{$-11.17$}} & \textcolor{blue}{\bm{$-49.43$}} & \textcolor{blue}{\bm{$-10.89$}} & \textcolor{blue}{\bm{$-47.36$}} & \textcolor{blue}{\bm{$-9.90$}} & \textcolor{blue}{\bm{$-50.84$}} \\\midrule
  \multicolumn{1}{c|}{MLIC$^+$ (Ours)}                                                & \textcolor{red}{\bm{$-11.39$}}      &\textcolor{red}{\bm{$-52.75$}}       &\textcolor{red}{\bm{$-16.38$}}  & \textcolor{red}{\bm{$-53.54$}} & \textcolor{red}{\bm{$-12.56$}} & \textcolor{red}{\bm{$-48.75$}} & \textcolor{red}{\bm{$-15.03$}} & \textcolor{red}{\bm{$-52.30$}} & \textcolor{red}{\bm{$-14.85$}} & \textcolor{red}{\bm{$-50.31$}}& \textcolor{red}{\bm{$-13.42$}} & \textcolor{red}{\bm{$-53.38$}}\\\midrule
  \end{tabular}
  \caption{BD-Rate $(\%)$ comparison for PSNR (dB) and MS-SSIM (dB), with the best ones in \textcolor{red}{red} and second-best ones in \textcolor{blue}{blue}. “$--$” means the result is not available. The anchor is VTM-17.0 Intra.}
  \label{tab:rd}
\end{table*}
\begin{figure*}
  \centering
  \includegraphics[width=\linewidth]
  {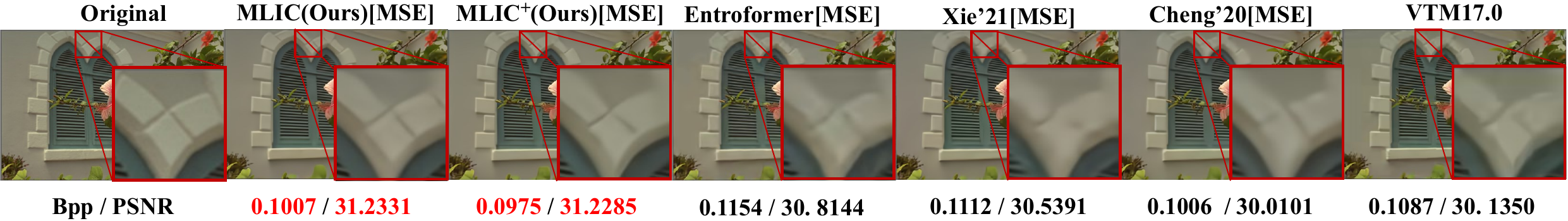}
  \caption{Visualization of the reconstructed Kodim07 from the Kodak dataset. The metrics are [bpp↓/PNSR↑].
  We compare our MLIC and MLIC$^+$ with Cheng'20~\cite{DBLP:conf/cvpr/ChengSTK20}, Xie'21~\cite{DBLP:conf/mm/XieCC21},
  Entroformer~\cite{DBLP:journals/corr/abs-2202-05492} and VTM-17.0~\cite{vtm2019}.}
  \label{visual}
\end{figure*}
\subsection{Performance}
We evaluate our models on rate-distortion performance and codec efficiency.
\begin{figure*}
  \centering
  \includegraphics[width=\linewidth]
  {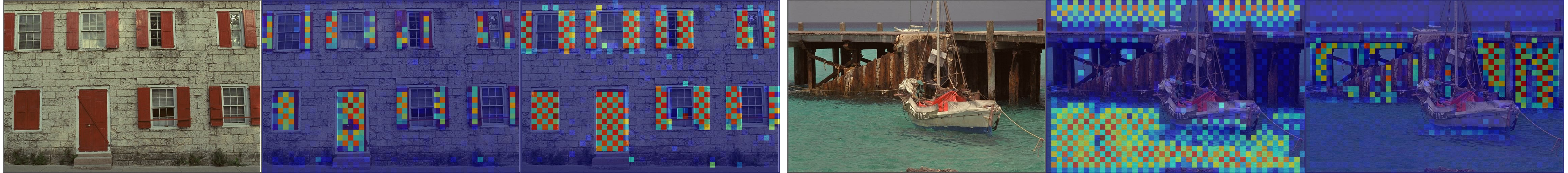}
  \caption{Attention map of Kodim01 and Kodim11 extracted by Intra-Global Context Model of MLIC (optimized for MSE, $\lambda=0.0035$).
  Because interactions within anchor and non-anchor part are not allowed,
  the attention map is checkerboard-like.}
  \label{kodim01_attn}
  \end{figure*}
\begin{table}[t]
  \begin{threeparttable}
  \centering
  \footnotesize
  \resizebox{\columnwidth}{!}{
  \begin{tabular}{@{}c|cccc }
  \toprule
  \multicolumn{2}{c|}{\multirow{2}{*}{Methods}}                       & \multicolumn{2}{c}{Kodak~\cite{kodak}}      \\
  \multicolumn{2}{c|}{}                                               & \multicolumn{1}{c}{Encoding Time ($\mathit{s}$)} & \multicolumn{1}{c}{Decoding Time ($\mathit{s}$)} \\ \midrule
  \multicolumn{2}{c|}{VTM-17.0~\cite{vtm2019}}                              & $104.9218$    & $0.2354$   \\\midrule
  \multicolumn{2}{c|}{Cheng'20 (CVPR'20)~\cite{DBLP:conf/cvpr/ChengSTK20}}            & $3.7082$      & $8.6586$    \\\midrule
  \multicolumn{2}{c|}{Minnen'20 (ICIP'20)~\cite{DBLP:conf/icip/MinnenS20}}            & $0.2467$      & $0.1298$    \\\midrule
  \multicolumn{2}{c|}{Xie'21 (MM'21)~\cite{DBLP:conf/mm/XieCC21}}                   & $4.0973$      & $9.1609$    \\\midrule
  \multicolumn{2}{c|}{Entroformer (ICLR'22)~\cite{DBLP:journals/corr/abs-2202-05492}} & $4.7682$      & $85.9190$   \\\midrule
  \multicolumn{2}{c|}{WACNN (CVPR'22)~\cite{DBLP:journals/corr/abs-2203-08450}}       & $0.2400$      & $0.1400$    \\\midrule
  \multicolumn{2}{c|}{STF (CVPR'22)~\cite{DBLP:journals/corr/abs-2203-08450}}         & $0.2594$      & $0.1629$    \\\midrule
  \multicolumn{2}{c|}{ELIC$^\ast$(CVPR'22)~\cite{He_2022_CVPR}}                            & $0.2315$           & $0.2057$         \\\midrule
  \multicolumn{2}{c|}{MLIC (Ours)}                                          & $0.2202$  & $0.1699$  \\\midrule
  \multicolumn{2}{c|}{MLIC$^+$ (Ours)}                                      & $0.3095$  & $0.2767$ \\\midrule
  \end{tabular}
  }
  \begin{tablenotes}
  \item\footnotesize{ELIC$^\ast$ is reimplemented by us because official ELIC is not open-sourced.}
  \end{tablenotes}
  \end{threeparttable}
  \caption{Encoding time and decoding time results compared with recent works.}
  \label{Tab:complex}
  \end{table}
\subsubsection{Rate-Distortion Performance}
Figure~\ref{rd} shows the rate-distortion performance on Kodak~\cite{kodak} dataset.
We report bd-rate reduction in Table~\ref{tab:rd} on Kodak~\cite{kodak},
Tecnick~\cite{asuni2014testimages}, CLIC Pro Val~\cite{clic2020dataset},
CLIC'21 Test~\cite{clic2021dataset}, CLIC'22 Test~\cite{clic2022dataset},
and JPEGAI Test~\cite{jpegai} datasets.
We compare our MLIC and MLIC$^+$ with recent
models~\cite{DBLP:conf/cvpr/ChengSTK20, DBLP:conf/icip/MinnenS20,
DBLP:conf/iclr/QianTSLLSHJ21,DBLP:conf/mm/XieCC21,
DBLP:journals/corr/abs-2202-05492,DBLP:journals/corr/abs-2203-08450,
He_2022_CVPR,Wang_2022_CVPR,DBLP:journals/corr/abs-2203-10897, zhu2021transformer,
DBLP:journals/corr/abs-2112-04487, pan2022content, koyuncu2022contextformer}
and VTM-17.0~\cite{vtm2019}.
Our MLIC and MLIC$^+$ achieve state-of-the-art performance on these datasets when measured
in PSNR and MS-SSIM.
Our MLIC and MLIC$^+$ reduce
BD-rate by $8.05\%$ and $11.39\%$ on Kodak dataset over VVC when measured in PSNR.
Compared with Cheng'20~\cite{DBLP:conf/cvpr/ChengSTK20},
our MLIC can achieve a maximum improvement of $0.5\sim0.8$dB in PSNR
and achieve a maximum improvement of $0.6$dB in MS-SSIM on Kodak,
our MLIC$^+$ can achieve a maximum improvement of $0.8\sim1.0$dB in PSNR.
Our MLIC and MLIC$^+$ adopt simplified analysis transform and synthesis transform of
Cheng'20~\cite{DBLP:conf/cvpr/ChengSTK20}, therefore,
the improvement of model performance is attributed to our Multi-Reference Entropy Models.
Our Multi-Reference Entropy Models can capture more contexts.
We illustrate the amount of contexts captured by different entropy models
in our supplementary material.
The improvement also proves correlations exist in multiple dimensions since
Cheng'20~\cite{DBLP:conf/cvpr/ChengSTK20} adopts an spatial autoregressive context model.
Compared with ELIC~\cite{He_2022_CVPR}, our MLIC$^+$ can be up to $0.4$db higher
at low bit rate and reduce BD-rate by $6.23\%$ over ELIC~\cite{He_2022_CVPR}.
\subsubsection{Qualitative Results}
    Figure~\ref{visual} illustrates the example of reconstructed Kodim07 of our MLIC,
    our MLIC$^+$, Entroformer~\cite{DBLP:journals/corr/abs-2202-05492},
    Xie'21~\cite{DBLP:conf/mm/XieCC21}, Cheng'20~\cite{DBLP:conf/cvpr/ChengSTK20}
    and VTM-17.0~\cite{vtm2019}. PSNR value of our reconstructed images are $1$dB
    higher than image reconstructed by VTM-17.0. Our reconstructed images retain
    more details with lower bpp. In terms of visual quality, our MLIC and MLIC$^+$
    have significant improvements compared to other models. We provide more qualitative
    results in our supplementary material.
\subsubsection{Codec Efficiency Analysis}
In MLIC and MLIC$^+$, Our local spatial and global spatial context models are parallel.
Although we divide $\hat {\boldsymbol y}$ into slices,
since MLIC has only $6$ slices and MLIC$^+$ has
only $10$ slices and the resolution of each slice is small,
the serial processing between slices does not add too much time.
We compare our MLIC and MLIC$^+$ with
other recent models~\cite{DBLP:conf/cvpr/ChengSTK20,DBLP:conf/icip/MinnenS20,
DBLP:conf/mm/XieCC21,DBLP:journals/corr/abs-2202-05492,
DBLP:journals/corr/abs-2203-08450,He_2022_CVPR}
on encoding time, and decoding time.
We include the arithmetic coding time.
We compare encoding and decoding time on Kodak~\cite{kodak}.
Our MLIC can encode and decode quite fast when compared with other models.
Slice the entropy model of MLIC$^+$ is more complex, it takes slightly longer time to
encode and decode an image.
\subsection{Ablation Studies}
\subsubsection{Settings}
We conduct corresponding ablation studies and
evaluate the contributions of proposed entropy models on
Kodak~\cite{kodak}.
Each model is optimized for MSE.
We train each model for $1.2$M steps.
We set learning rate to $10^{-4}$ and batch size to $8$. We crop images to $256\times 256$
patches during ablation studies.
The results and configure are shown in Table~\ref{tab:ablation}.
The base model is MLIC \textit{w/o} context modules.
\subsubsection{Analysis of Channel-wise Context Module}
Channel-wise context module leads to a significant improvement in performance,
possible to refer to symbols in the same and close position in the previous slices.
The effectiveness of channel-wise context module proves the redundancy among channels.
\begin{table}
  \centering
  \footnotesize
  \begin{tabular}{cccc}
  \toprule
               & Kodak~\cite{kodak}    \\ \midrule
          VTM-17.0~\cite{vtm2019} & $0.000$ \\\midrule
          base + $g_{lc,ckbd}$ & $+14.82$ \\\midrule
          base + $g_{lc,stk}$ & $+13.01$ \\\midrule
          base + $g_{lc,attn}$ & $+9.95$ \\\midrule
          base + $g_{ch}$ & $+2.73$ \\\midrule
          base + $g_{lc,ckbd} + g_{ch}$   & $-0.98$ \\\midrule
          base + $g_{lc,stk} + g_{ch}$   & $-2.23$ \\\midrule
          base + $g_{lc,attn} + g_{ch}$   & $-2.92$ \\\midrule
          base + $g_{ch} + g_{gc,intra}$   & $+1.24$ \\\midrule
          base + $g_{ch} + g_{gc,intra, w/o\enspace mask}$   & $+1.91$ \\\midrule
          base + $g_{lc,stk} + g_{ch} + g_{gc,intra}$ & $-4.11$ \\\midrule
          base + $g_{lc,attn} + g_{ch} + g_{gc,intra}$ & $-4.90$  \\\midrule
          base + $g_{lc,attn} + g_{ch} + g_{gc,intra} + g_{gc,inter}$ & $-5.63$ \\\midrule
  \end{tabular}
  \caption{Ablation studies on Kodak~\cite{kodak}.
  The metric is BD-Rate (\%) for PSNR (dB). The anchor is VTM-17.0 Intra.
  }
  \label{tab:ablation}
\end{table}
\subsubsection{Analysis of Local Context Module}
Vanilla checkerboard context module leads to slight performance degradation while
allows for two-pass decoding.
Stacked checkerboard context module increases the depth for non-linearity, which
leads to more powerful expressiveness. In ablation studies, $g_{lc, stk}$
saves $1.81\%$ more bit-rates compared to $g_{lc, ckbd}$.
Checkerboard attention module performs much better. $g_{lc, attn}$
saves $4.87\%$ more bit-rates compared to $g_{lc, ckbd}$,
which can be attributed to the context-adaption and non-linearity of our
proposed $g_{lc, attn}$.
\subsubsection{Analysis of Global Context Module}
We illustrate attention map of Intra-Slice Context Module in Figure~\ref{intra_attn}.
Our model successfully captures distant correlations,
which are impossible for local context models to capture.
Our Intra-Slice Global Context Module may be somewhat similar to the cross-attention model.
However, we don't care about interactions between these two slices.
We only use the attention map of $\hat {\boldsymbol{y}}^{i-1}$ to predict correlations in $\hat {\boldsymbol{y}}^i$.
We also remove the mask in Intra-Slice Global Context Module $g_{gc, intra}$.
Removing mask leads to performance degradation, because removing mask
makes it hard for network to learn.
When our proposed global context modules cooperate with local social context modules,
the performance is further improved, which proves the necessarity of
global spatial context modules for global correlation capturing and local spatial
context modules for local correlation capturing.
The gain of Inter-Slice Global Context Module is not very huge,
which can be attributed to the approximation via anchor parts.
One problem of our methods for global context is their quadratic computational complexity.
One solution is cropping an image into patches.
We find shared attention map and cropping an image into non-overlapped patches has almost no influence on performance.
The results are reported in our supplementary material.
\section{Conclusion}\label{Sec:conclusion}
In this paper, we propose multi-reference entropy models MEM and MEM$^+$,
which capture correlations in multiple dimensions. To our knowledge,
this is the first successful attempt to capture channel, local and global correlations.
Based on MEM and MEM$^+$, we obtain state-of-the-art models MLIC and MLIC$^+$.
The significance of our work is investigating multiple correlations
in latent representation and exploring the potential of an entropy model.
However, due to the computational overhead, MLIC and MLIC$^+$ cannot be directly
applied to mobile devices. We expect that this problem can be addressed by
using knowledge distillation, network pruning, structural re-parameterization,
and other light-weight designs.
\begin{acks}
We thank the anonymous reviewers for their valuable comments and suggestions.
This work is financially supported by National Natural Science Foundation of China U21B2012 and  62072013, Shenzhen Science and Technology Program-Shenzhen Cultivation of Excellent Scientific and Technological Innovation Talents project(Grant No. RCJC20200714114435057) , Shenzhen Science and Technology Program-Shenzhen Hong Kong joint funding project (Grant No. SGDX20211123144400001), this work is also financially supported for Outstanding Talents Training Fund in Shenzhen.
\end{acks}

\appendix
\section{More Netwrok Architecture Details}
\begin{figure}
  \centering
  \includegraphics[width=0.8\linewidth]
  {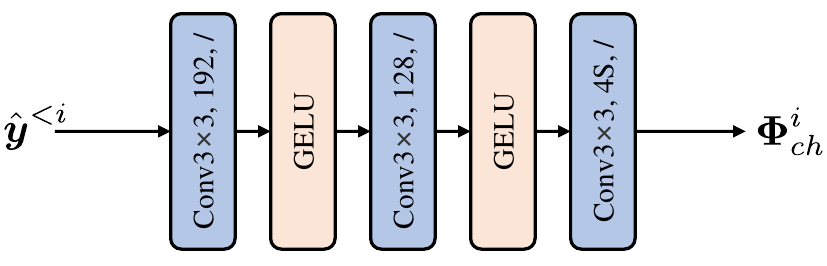}
  \caption{Channel-wise Context Module $g_{ch}$.}
  \label{fig:gch}
\end{figure}

\begin{figure}
  \centering
  \includegraphics[width=0.8\linewidth]
  {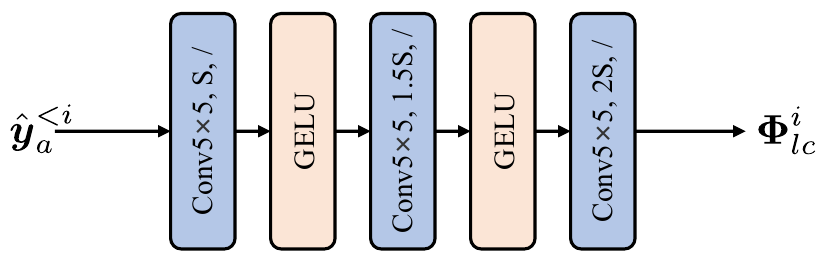}
  \caption{Stacked Checkerboard Context Module $g_{lc, stk}$.}
  \label{fig:stk}
  \end{figure}

\begin{figure}
  \centering
  \includegraphics[width=0.8\linewidth]
  {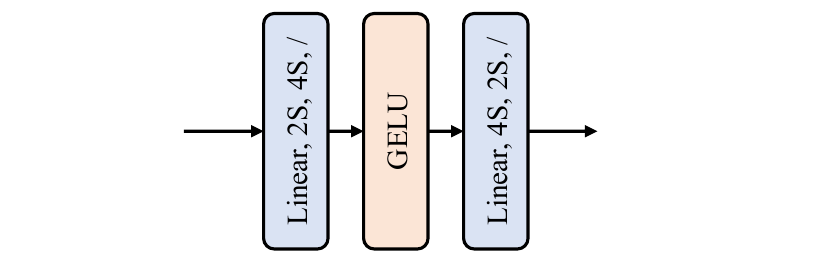}
  \caption{Feed Forward Network (FFN).}
  \label{fig:ffn}
  \end{figure}

\begin{figure}
  \centering
  \includegraphics[width=\linewidth]
  {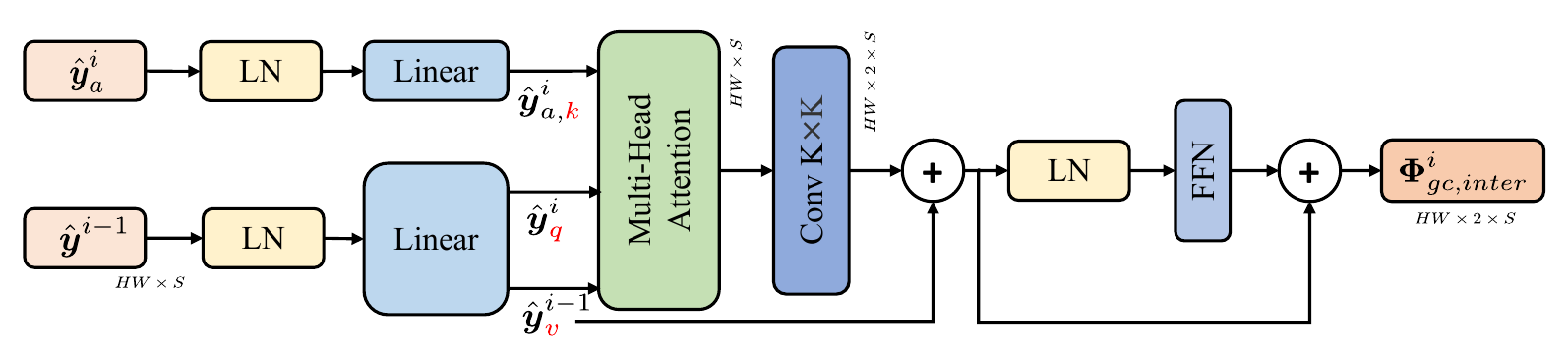}
  \caption{Inter-Slice Global Context Module $g_{gc, inter}$. $S$ is the channel number of a slice.}
  \label{fig:inter_arch}
  \end{figure}

When encoding, we split latent representation ${\hat {\boldsymbol y}}$
into anchor part ${\hat {\boldsymbol y}}^i_a$ and nonanchor pact
${\hat {\boldsymbol y}}^i_{na}$. The architecture of our
Channel Context Module is illustrated in Figure~\ref{fig:gch}.
The architecture of Stacked Checkerboard Context Module is illustrated in Figure~\ref{fig:stk},
which contains three $5\times 5$ convolutional layers.
We use GELU~\cite{hendrycks2016gaussian} for non-linearity.
The architecture of Stacked Checkerboard Context Module is
illustrated in Figure~\ref{fig:ckbd_attn_arch}.
The architecture of FFN is illustrated in Figure~\ref{fig:ffn}.
The architecture of inter-slice global context model is illustrated in Figure~\ref{fig:inter_arch}.
\begin{figure}[t!]
	\centering
	\subfloat{
		\includegraphics[scale=0.4]{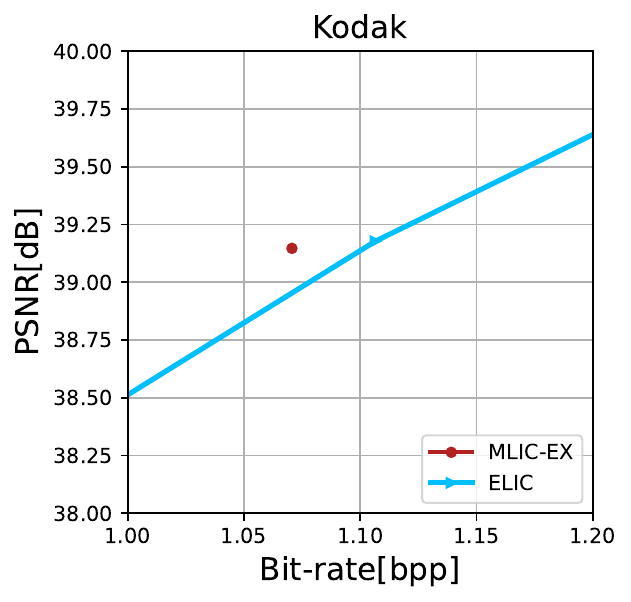}}
	\subfloat{
		\includegraphics[scale=0.4]{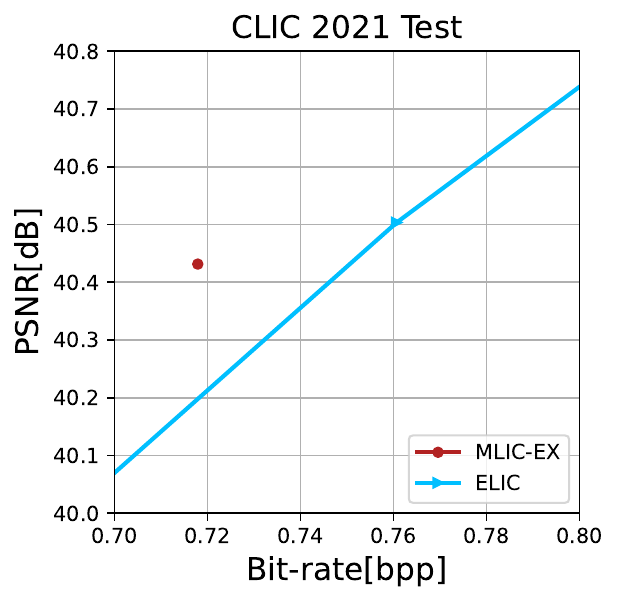}}
	\caption{PSNR-Bit-rate curve on Kodak and CLIC 2021 Test dataset.}
	\label{fig:mlicexrd}
\end{figure}
  \begin{figure*}
  \centering
  \includegraphics[width=0.9\linewidth]
  {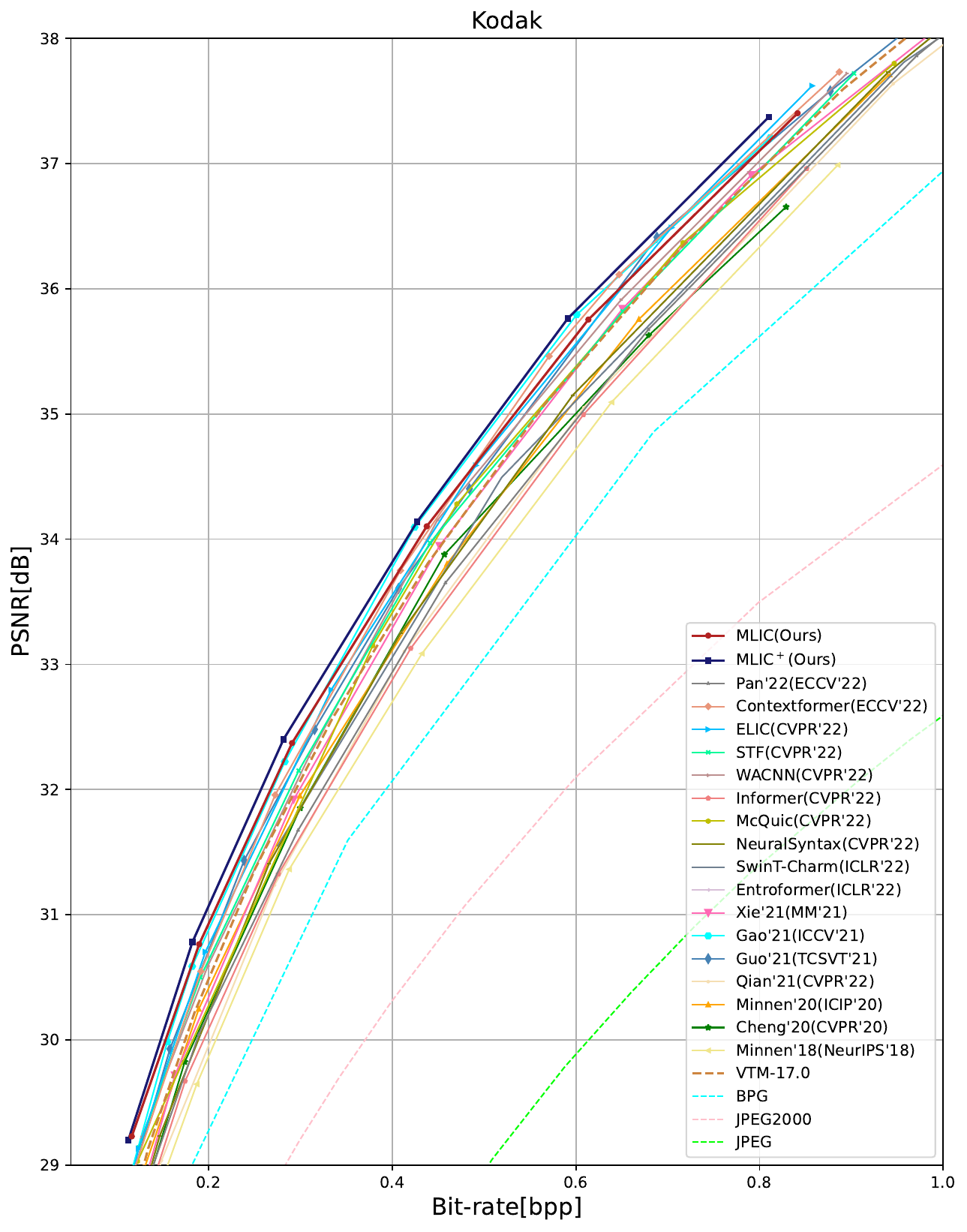}
  \caption{Rate-distortion data of MLIC and MLIC+ on Kodak dataset, which contains $24$ raw images.}
  \label{fig:kodak_psnr_large}
  \end{figure*}
  \begin{figure*}
  \centering
  \includegraphics[width=0.9\linewidth]
  {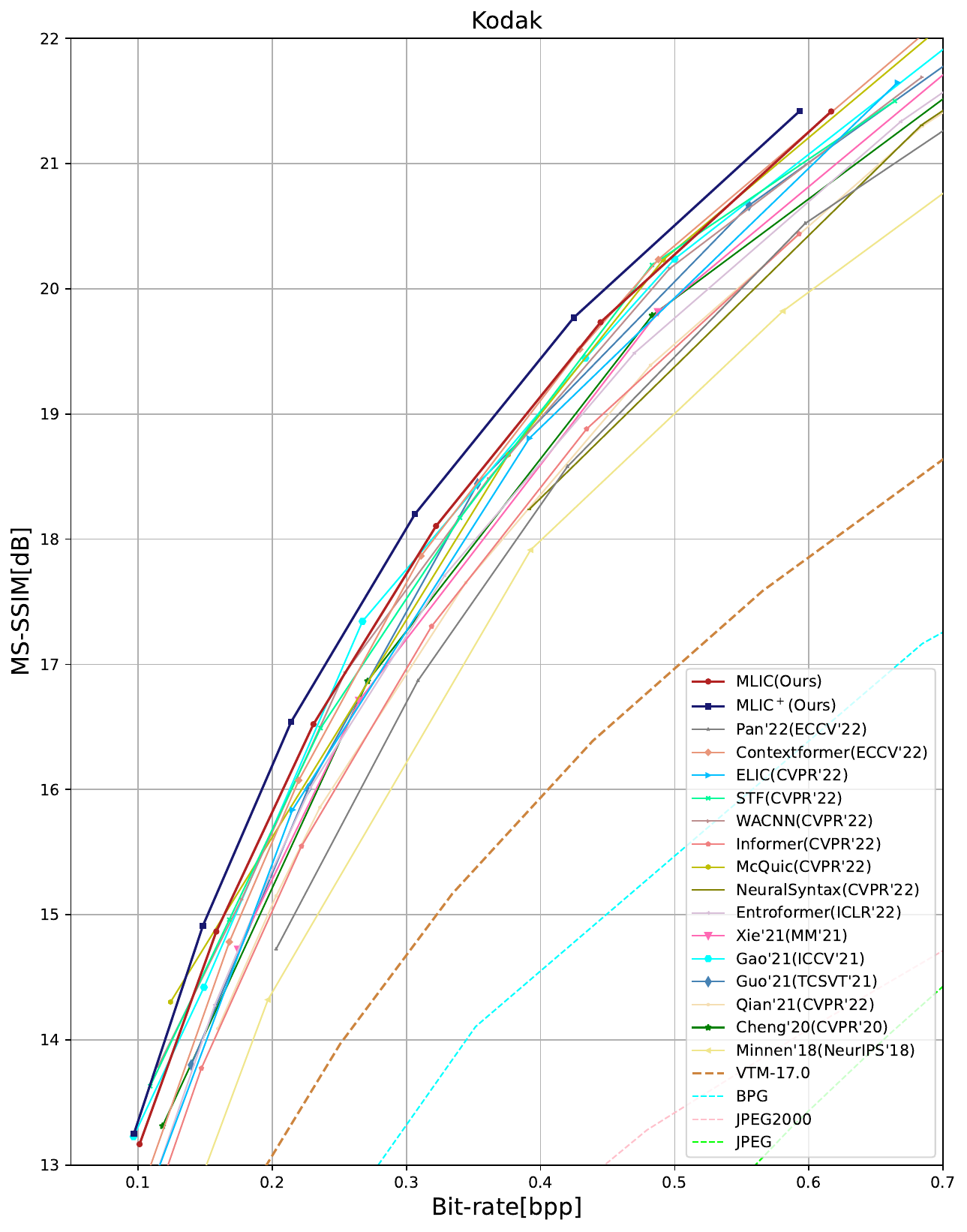}
  \caption{Rate-distortion data of MLIC on Kodak dataset, which contains $24$ raw images.}
  \label{fig:kodak_msssim_large}
  \end{figure*}
  \begin{figure*}
  \centering
  \includegraphics[width=0.9\linewidth]
  {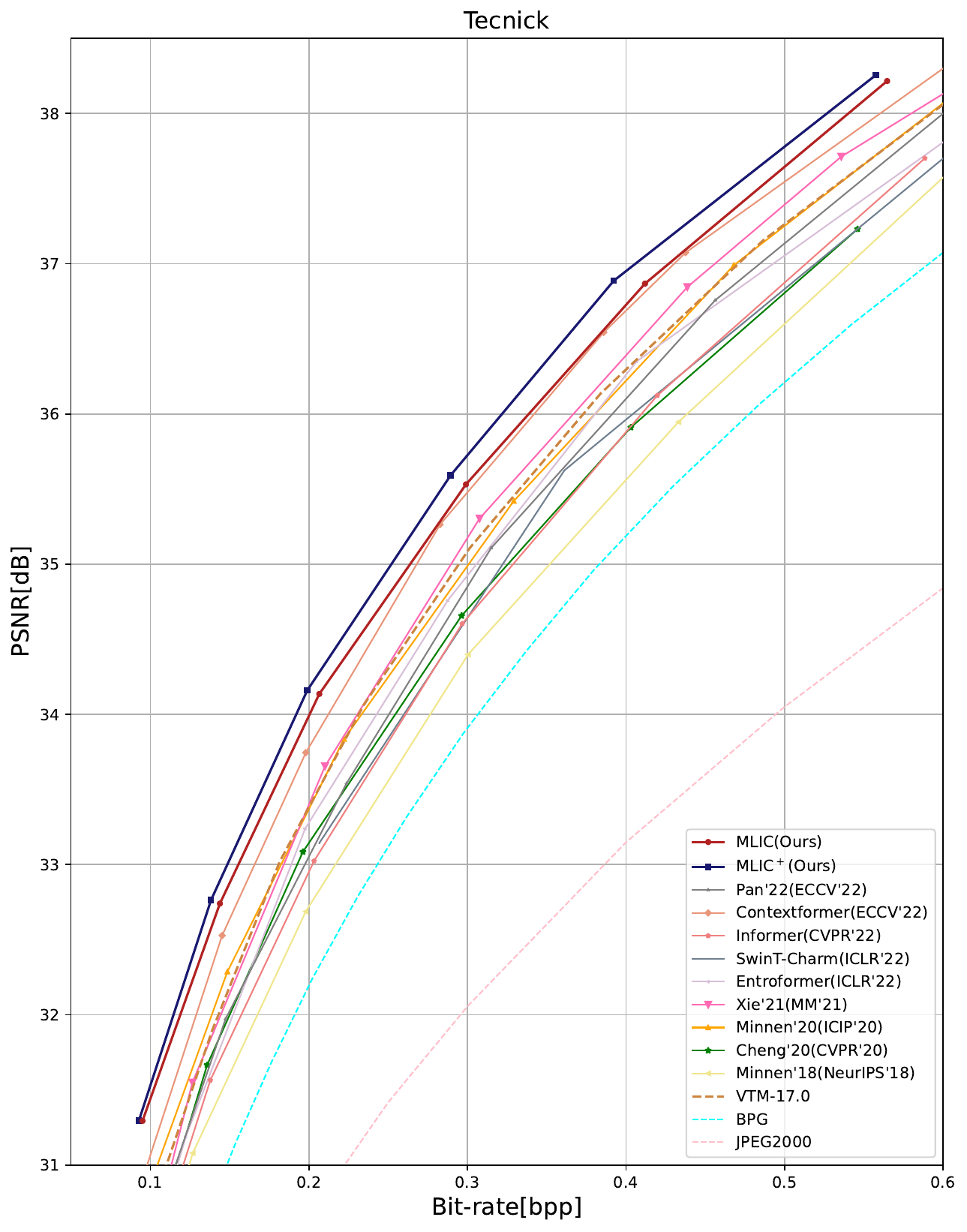}
  \caption{Rate-distortion data of MLIC and MLIC+ on Tecnick dataset, which contains $100$ raw images. All images are padded to multiples of $64$.}
  \label{fig:tecnick_psnr_large}
  \end{figure*}
  \begin{figure*}
  \centering
  \includegraphics[width=0.9\linewidth]
  {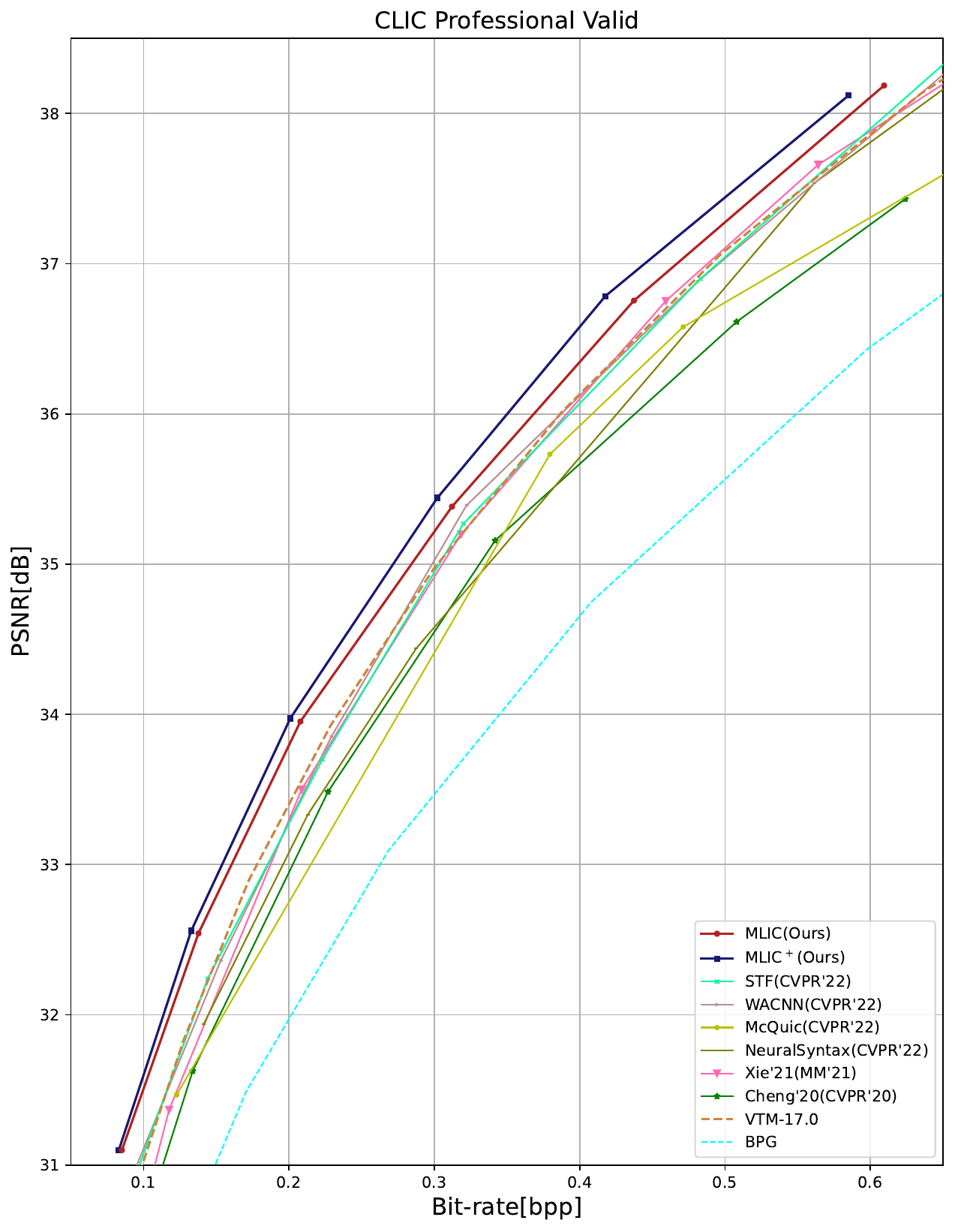}
  \caption{Rate-distortion data of MLIC and MLIC+ on CLIC Professional Valid dataset, which contains $41$ raw images. All images are padded to multiples of $64$.}
  \label{fig:clicprofessional_psnr_large}
  \end{figure*}
  \begin{figure*}
  \centering
  \includegraphics[width=0.9\linewidth]
  {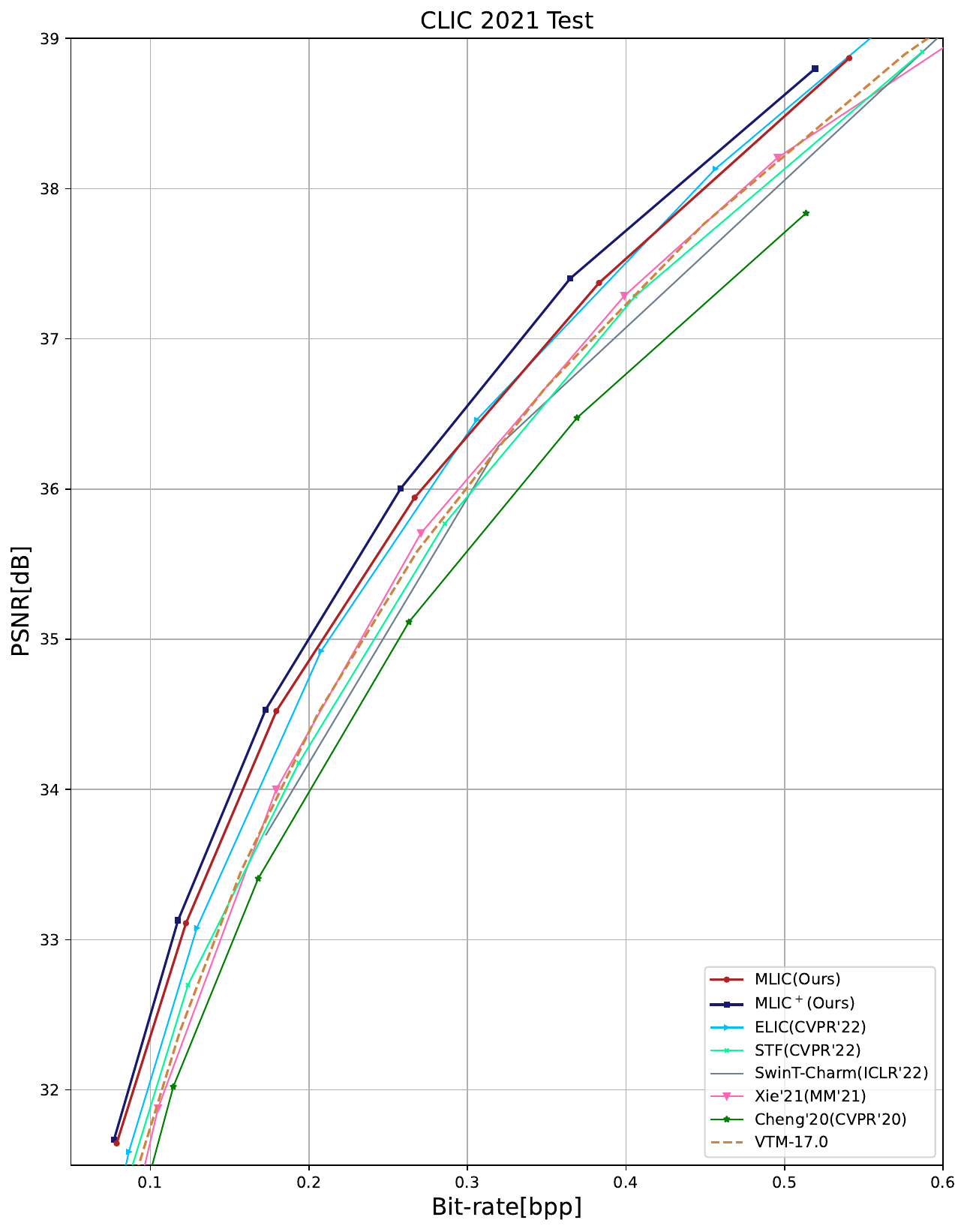}
  \caption{Rate-distortion data of MLIC and MLIC+ on CLIC 2021 Test dataset, which contains $60$ raw images. All images are padded to multiples of $64$.}
  \label{fig:clic2021test_psnr_large}
  \end{figure*}
  \begin{figure*}
    \centering
    \includegraphics[width=0.9\linewidth]
    {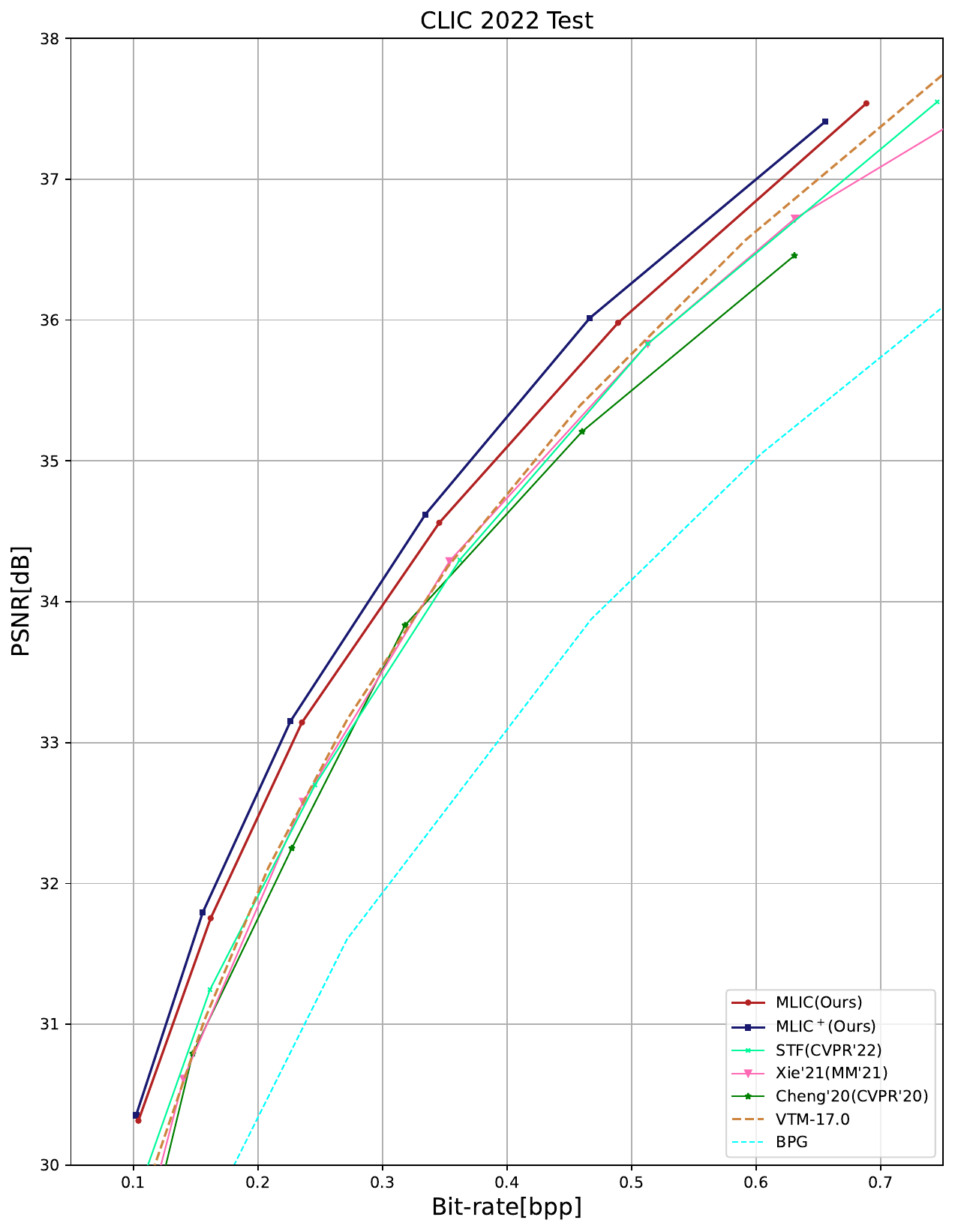}
    \caption{Rate-distortion data of MLIC and MLIC+ on CLIC 2021 Test dataset, which contains $30$ raw images. All images are padded to multiples of $64$.}
    \label{fig:clic2022test_psnr_large}
    \end{figure*}
  \begin{figure*}
  \centering
  \includegraphics[width=0.9\linewidth]
  {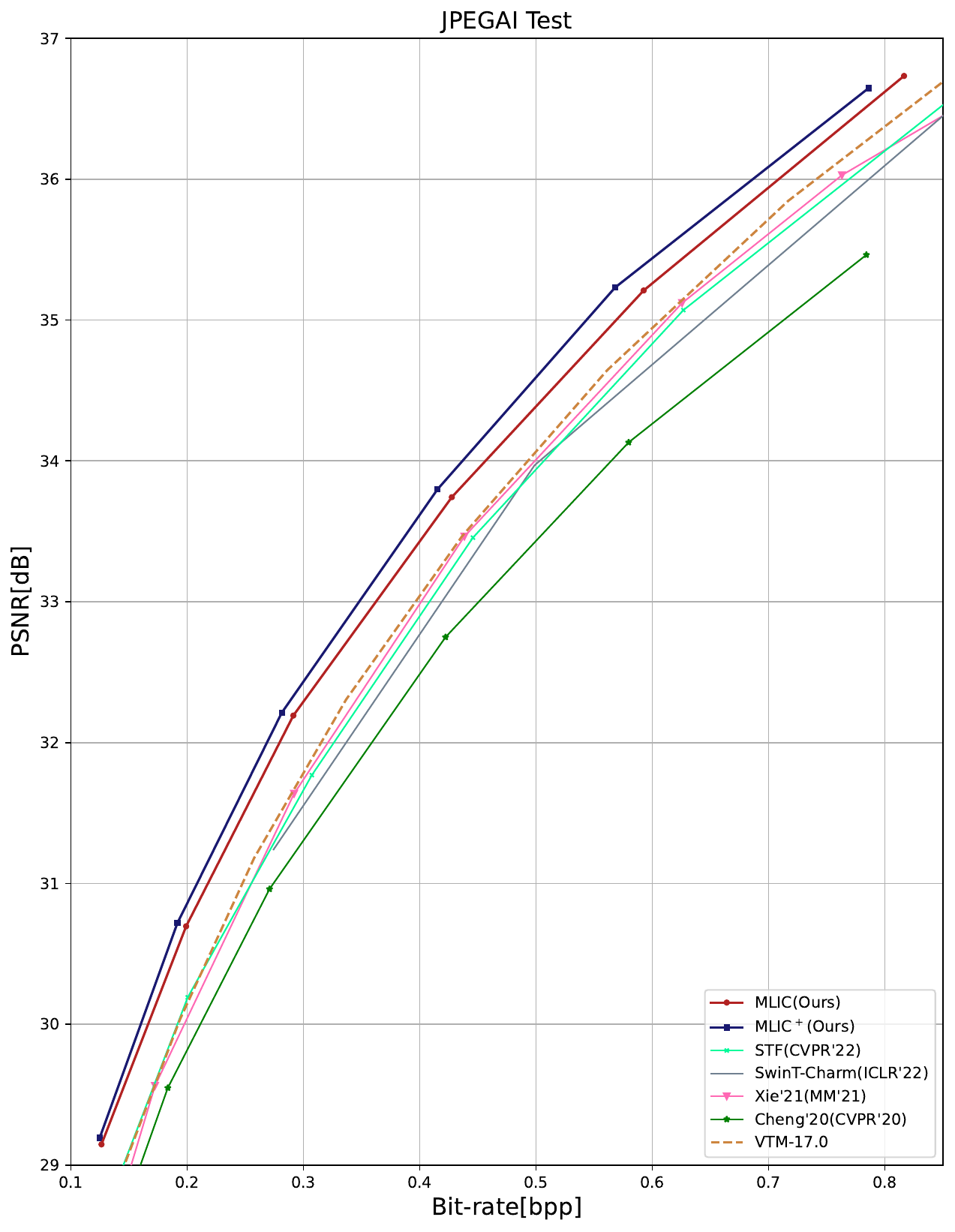}
  \caption{Rate-distortion data of MLIC and MLIC+ on JPEGAI Test dataset, which contains $16$ raw images. All images are padded to multiples of $64$.}
  \label{fig:jpegaitest_psnr_large}
  \end{figure*}
\section{Detailed Experiment Settings}
We implement our MLIC and MLIC+ on Pytorch 1.10.0~\cite{paszke2019pytorch},
CompressAI 1.2.0b3~\cite{DBLP:journals/corr/abs-2011-03029} and Python 3.9.7.
We use a GeForce RTX 3090 GPU and an Xeon Silver 4210R on Ubuntu 20.04
to test encoding and decoding latency.
We enable the deterministic inference mode when testing the model speeds.
\section{Experiments on Other Backbone and Simpler Global Context Module}
As stated in the main paper, using Cheng'20~\cite{DBLP:conf/cvpr/ChengSTK20}
without attention modules leads to performance degradation at high bit-rates.
We replace it with backbone of ELIC~\cite{He_2022_CVPR}. Different from ELIC,
we evenly divide the latent representation into slices.
We simplify the Intra-Slice Global Context Module by sharing attention maps.
We use the attention map of the first slice to predict global correlations in other slices.
The process of Intra-Slice Global Context Module can be formulated as:
\begin{equation}
    \hat {\boldsymbol{y}}^{i}_{attn} = \textrm{softmax}\left(\frac{{{\hat {\boldsymbol{y}}}^{1}}_{na,q} \times ({{\hat {\boldsymbol{y}}}^{1}}_{a,k})^\top}{\sqrt{S}} + mask\right) \times \hat {\boldsymbol{y}}^{i}_{a,v},
\end{equation}
\begin{equation}
    \hat {\boldsymbol{y}}^{i}_{conv} = \textrm{conv}_{K\times K}(\hat {\boldsymbol{y}}^{i}_{attn}) + \hat {\boldsymbol{y}}^{i}_{a,v},
\end{equation}
\begin{equation}
    {\boldsymbol{\Phi}}^i_{gc,intra} = \textrm{FFN}(\hat {\boldsymbol{y}}^{i}_{conv}) + \hat {\boldsymbol{y}}^{i}_{conv},
\end{equation}
where $\hat {\boldsymbol{y}}^{1}_{na,q}, \hat {\boldsymbol{y}}^{1}_{a,k} = \textrm{Embed}(\hat {\boldsymbol{y}}^{1})$, $\hat {\boldsymbol{y}}^{i}_{a,v} = \textrm{Embed}(\hat {\boldsymbol{y}}^i_{a})$.
We remove Inter-Slice Global Context Module and Latent Residual Prediction modules~\cite{DBLP:conf/icip/MinnenS20}
to further reduce complexity.
We call this model MLIC-EX, meaning extra version of MLIC.
We optimize MLIC-EX for MSE. We set $\lambda$ to $0.08$ to evaluate
the performance on high bit-rates. We compare our MLIC-EX on Kodak and CLIC 2021 Test dataset.
The performance of MLIC-EX is shown in Figure \ref{fig:mlicexrd}. Although
sharing attention maps leads to slight performance drop, our MLIC-EX outperforms ELIC.
Our MLIC-EX ranked {\bm{$3$}}-rd place at performance track of
VCIP'22 Practical End-to-End Image Compression Challenge at
\url{http://www.vcip2022.org/Challenge2.htm}.
\section{More Rate-Distortion Performance Results}
We illustrate detailed rate-distortion performance in Figure~\ref{fig:kodak_psnr_large}, \ref{fig:kodak_msssim_large},
 \ref{fig:tecnick_psnr_large}, \ref{fig:clicprofessional_psnr_large}, \ref{fig:clic2021test_psnr_large}, \ref{fig:jpegaitest_psnr_large}.\par
Kodak dataset is availavle at \url{http://r0k.us/graphics/kodak}.\par Tecnick dataset is availavle at \url{https://sourceforge.net/projects/testimages/files/OLD/OLD_SAMPLING/testimages.zip}.\par
CLIC Professional Valid dataset is available at \url{https://data.vision.ee.ethz.ch/cvl/clic/professional_valid_2020.zip}.\par
CLIC2021 Test dataset is available at \url{https://storage.googleapis.com/clic2021_public/professional_test_2021.zip}.\par
JPEGAI Test dataset is available at \url{https://jpegai.github.io/test_images/}.\par
We compare our methods with learned image compression methods Minnen'18 \cite{DBLP:conf/nips/MinnenBT18}, Cheng'20 \cite{DBLP:conf/cvpr/ChengSTK20}, Minnen'20 \cite{DBLP:conf/icip/MinnenS20}, Guo'21 \cite{ DBLP:journals/tcsv/GuoZFC22}, Gao'21 \cite{DBLP:conf/iccv/GaoYPHZDL21}, Xie'21 \cite{DBLP:conf/mm/XieCC21}, Entroformer \cite{DBLP:journals/corr/abs-2202-05492}, STF \cite{DBLP:journals/corr/abs-2203-08450},
ELIC \cite{He_2022_CVPR}, Contextformer \cite{koyuncu2022contextformer} and traditional image compression methods JPEG \cite{JPEG-ITU1992Information}, JPEG2000 \cite{DBLP:conf/icmcs/CharrierCL99}, BPG \cite{bpg}, VTM \cite{vtm2019}. \par
When evaluating performance on CLIC Professional Valid dataset \cite{clic2020dataset}, Tecnick \cite{asuni2014testimages} and CLIC2021 Test \cite{clic2020dataset}, we pad images to multiples of $64$.
There are few results on JPEGAI Test. We only compare our MLIC and MLIC$^+$ with Xie'21 \cite{DBLP:conf/mm/XieCC21}.
We obtain rate-distortion data of other models by asking authors via emails.
We get the rate-distortion performance of VTM-17.0 \cite{vtm2019} via following commands:
\begin{python}
  # Convert png image to yuv444 image
  ffmpeg -i $PNGINPUT
         -s $WIDTHx$HEIGHT
         -pix_fmt yuv444p $YUVOUTPUT
  # Encode
  ./EncoderApp -c cfg/encoder_intra_vtm.cfg
  -i $YUVOUTPUT -q $QP,
  -o /dev/null -b $BINOUTPUT
  --SourceWidth=$WIDTH
  --SourceHeight=$HEIGHT
  --FrameRate=1
  --FramesToBeEncoded=1
  --InputBitDepth=8
  --InputChromaFormat=444
  --ConformanceWindowMode=1
  # Decode
  ./DecoderApp -b $BINOUTPUT
  -o $RECYUVOUTPUT -d 8
  # Convert yuv444 image to png image
  ffmpeg -s $WIDTHx$HEIGHT
         -pix_fmt yuv444p
         -i $RECYUVOUTPUT $RECPNGOUTPUT
\end{python}
Since VTM-17.0~\cite{vtm2019} takes a long time to encode,
we provide the rate-distortion data of VTM-17.0 for other researchers to compare the performance.
\begin{table}[t]
  \centering
  \small
  \begin{tabular}{cccccc}
  \toprule
            Bpp    & PSNR    & MS-SSIM   \\ \midrule
            $0.03479003906250001$ & $25.386870719686915$ & $8.047509531502621$ \\\midrule
            $0.0487721761067708$ & $26.2384553268167622$ & $8.8334493567332508$ \\\midrule
            $0.09683820936414929$ & $28.11424989629592$ & $10.65856918948002$ \\\midrule
            $0.1341374715169271$ & $29.1327816260354062$ & $11.6864511628480159$ \\\midrule
            $0.1845033433702257$ & $30.2170780024719363$ & $12.8022351344119869$ \\\midrule
            $0.2506527370876736$ & $31.3689248084602177$ & $13.9678017798159022$ \\\midrule
            $0.3342785305447049$ & $32.5617004128156324$ & $15.1683148205680620$ \\\midrule
            $0.4387919108072917$ & $33.8079252181964094$ & $16.3874390222642639$ \\\midrule
            $0.5654881795247394$ & $35.0726027661156863$ & $17.5864643832630385$ \\\midrule
            $0.7151201036241318$ & $36.3319105006136311$ & $18.7555160888172701$ \\\midrule
            $0.8913472493489581$ & $37.5965940611501495$ & $19.9115141097121544$ \\\midrule
  \end{tabular}
  \caption{Rate-distortion data of VTM-17.0 on Kodak~\cite{kodak} dataset.}
  \label{tab:vtm_kodak}
\end{table}
\section{More Visualizations}
\subsection{Visualization of Reconstructed Images}
\begin{figure*}
  \centering
  \includegraphics[width=0.8\linewidth]
  {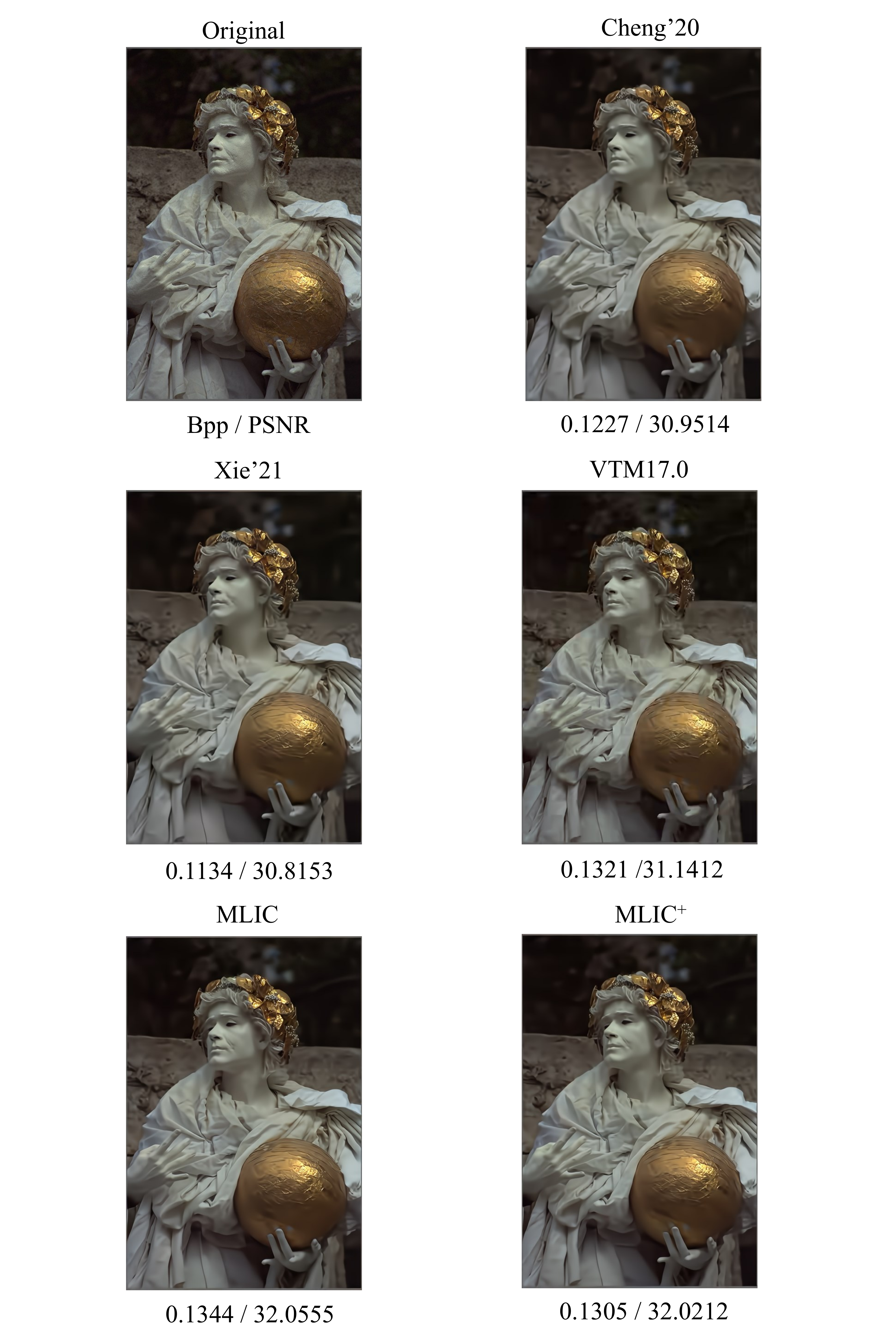}
  \caption{Qualitative comparison on reconstructed kodim17.  The metrics are [bpp↓/PNSR↑].}
  \label{fig:kodim17compare}
  \end{figure*}

\begin{figure*}
  \centering
  \includegraphics[width=0.8\linewidth]
  {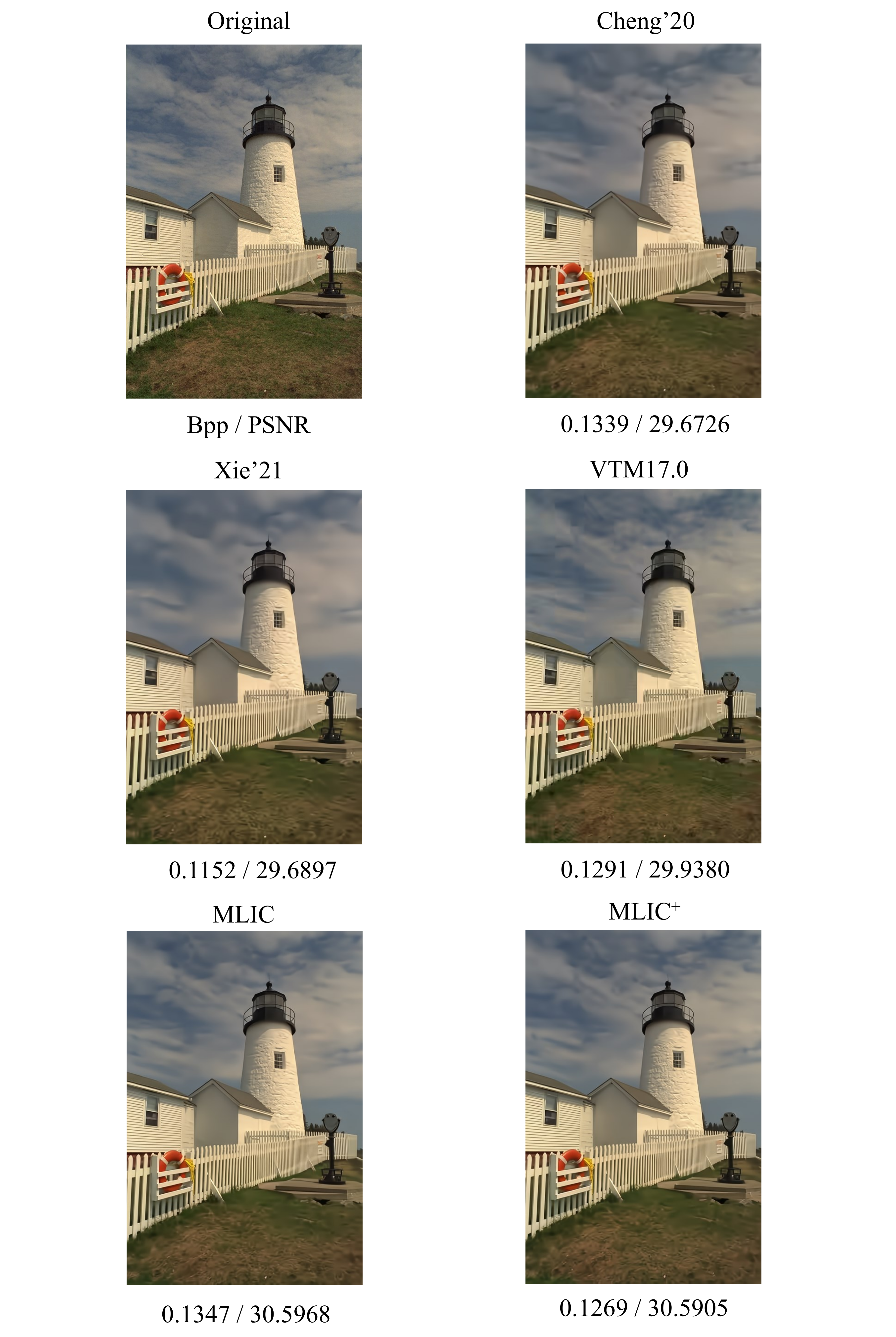}
  \caption{Qualitative comparison on reconstructed kodim19. The metrics are [bpp↓/PNSR↑].}
  \label{fig:kodim19compare}
  \end{figure*}

We select some reconstructed images from Kodak \cite{kodak}.
We compare our MLIC and MLIC+ with Xie'21~\cite{DBLP:conf/mm/XieCC21},
Cheng'20 \cite{DBLP:conf/cvpr/ChengSTK20} and VTM-17.0 \cite{vtm2019}.
Please see Figure \ref{fig:kodim17compare}, \ref{fig:kodim19compare} for reconstructed images.
\subsection{Visualization of Attention Maps}
\begin{figure*}
  \centering
  \includegraphics[width=0.85\linewidth]
  {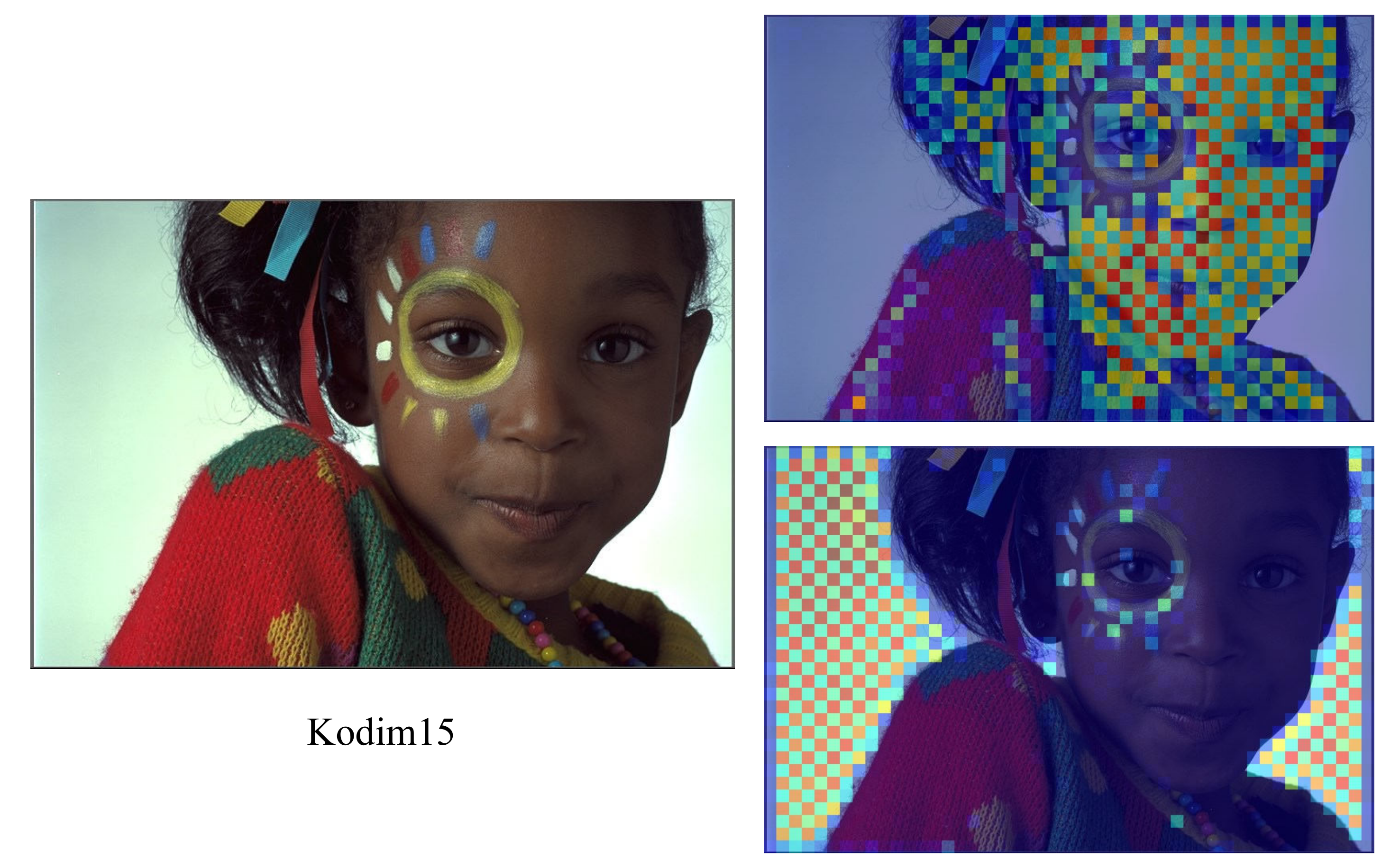}
  \caption{Attention map of Intra-Slice Global Context Model of MLIC(optimized for MSE, $\lambda=0.0035$). Because we divide latent representation into anchor and non-anchor, the attention map is checkerboard-like.}
  \label{fig:kodim15_attn_map}
  \end{figure*}

\begin{figure*}
  \centering
  \includegraphics[width=0.85\linewidth]
  {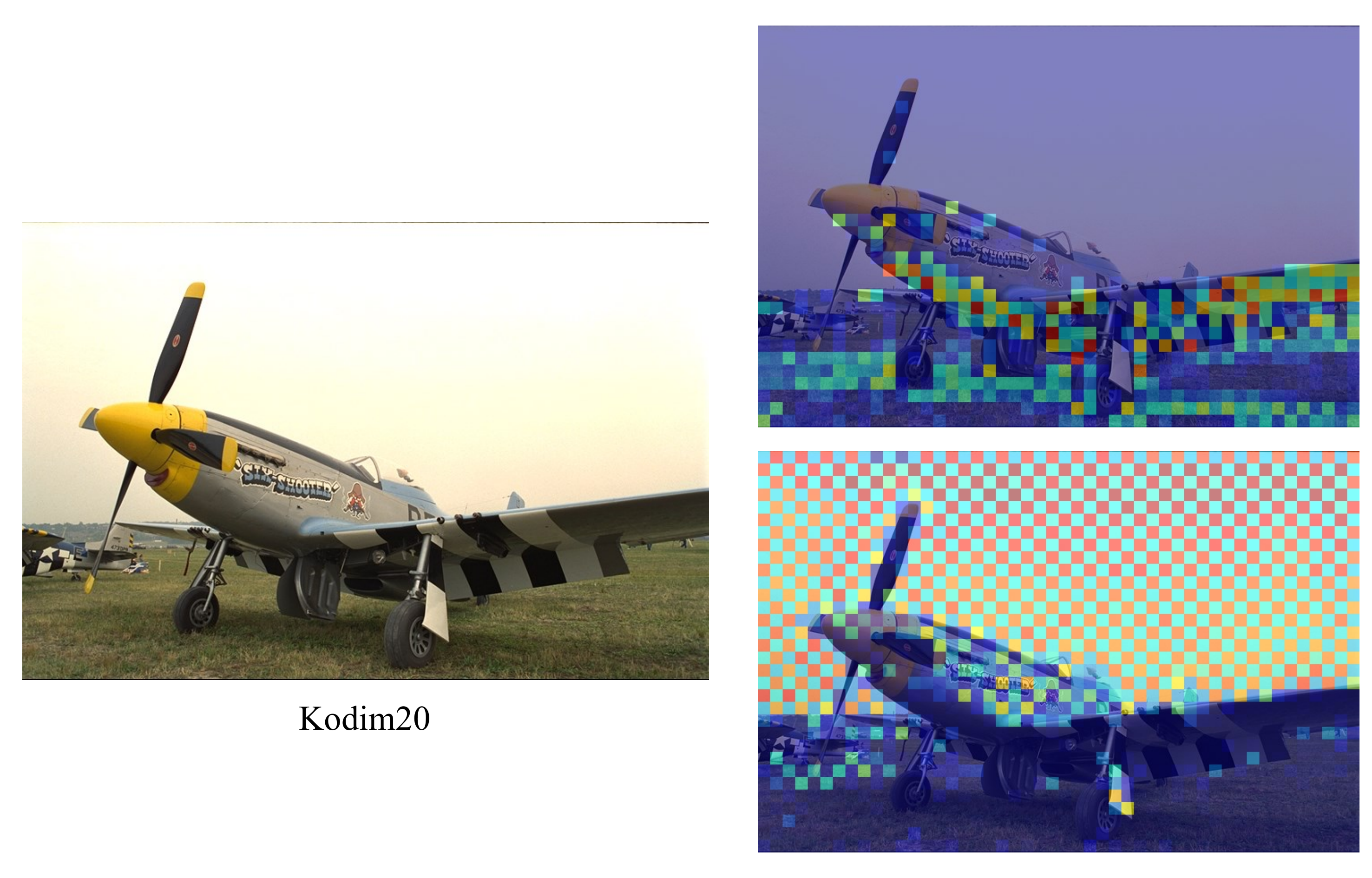}
  \caption{Attention map of Intra-Slice Global Context Model of MLIC(optimized for MSE, $\lambda=0.0035$). Because we divide latent representation into anchor and non-anchor, the attention map is checkerboard-like.}
  \label{fig:kodim20_attn_map}
  \end{figure*}

\begin{figure*}
  \centering
  \includegraphics[width=\linewidth]
  {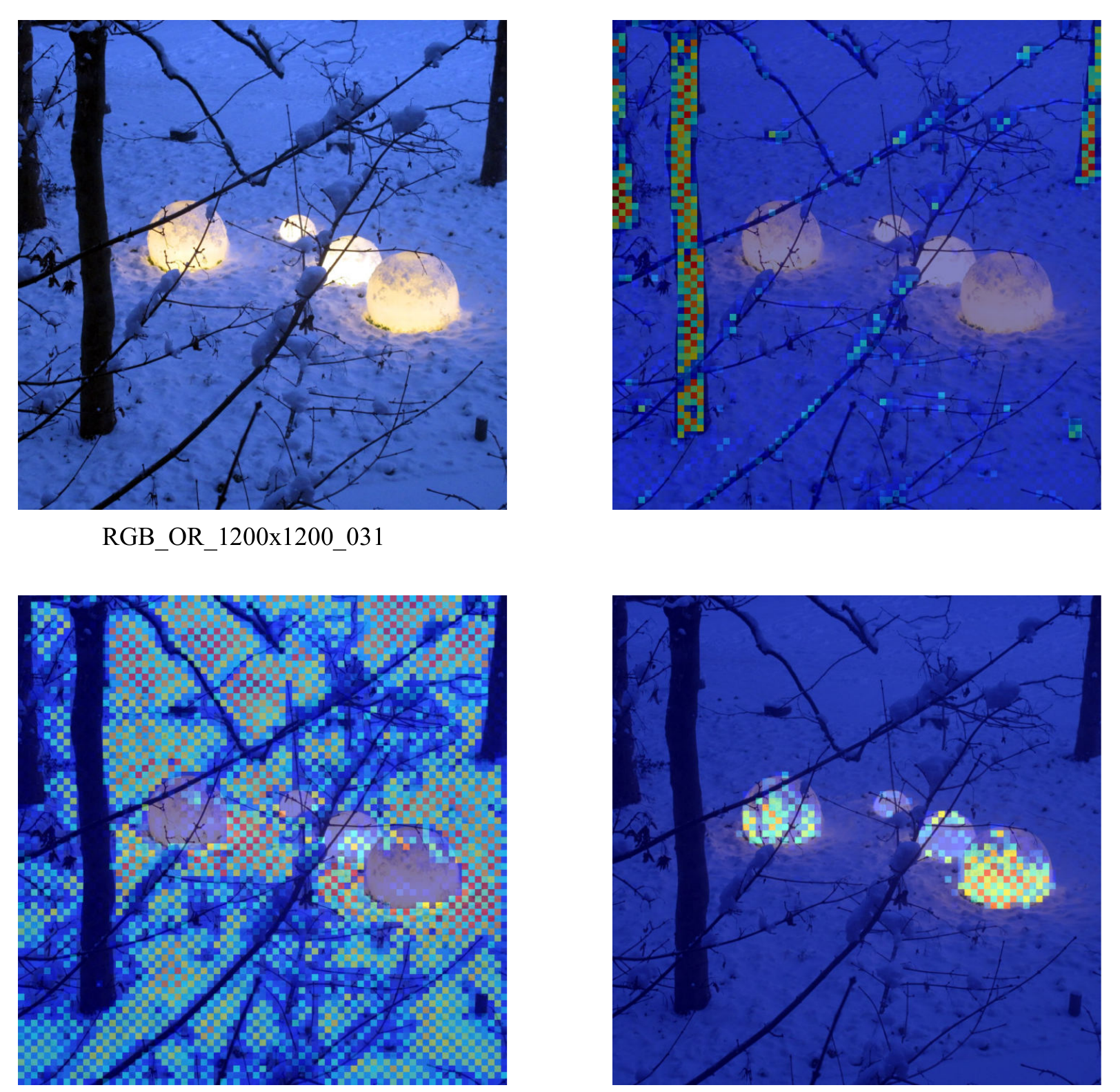}
  \caption{Attention map of Intra-Slice Global Context Model of MLIC(optimized for MSE, $\lambda=0.0035$). Because we divide latent representation into anchor and non-anchor, the attention map is checkerboard-like.}
  \label{fig:tecnick02_attn_map}
  \end{figure*}

\begin{figure*}
  \centering
  \includegraphics[width=\linewidth]
  {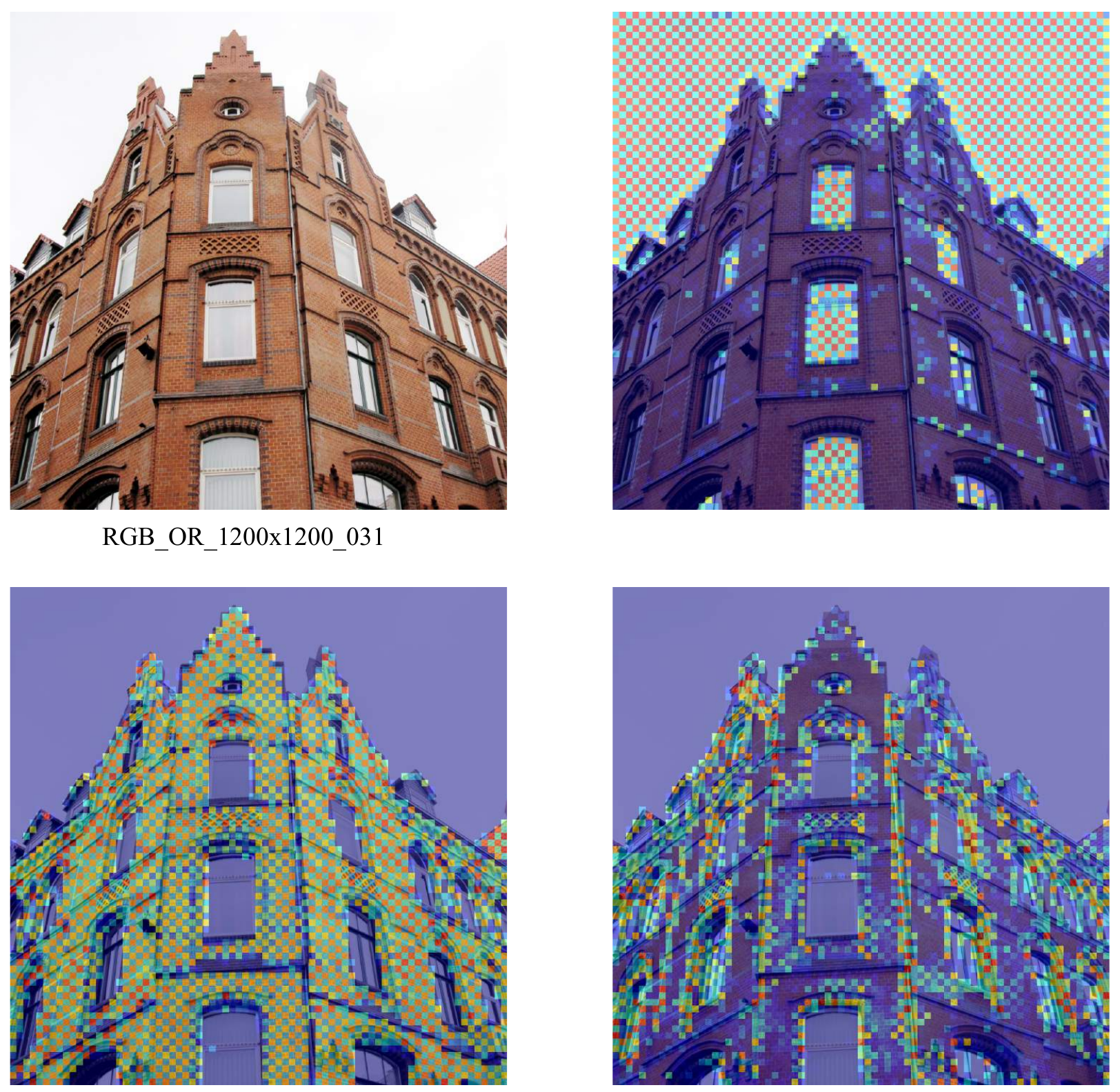}
  \caption{Attention map of Intra-Slice Global Context Model of MLIC(optimized for MSE, $\lambda=0.0035$). Because we divide latent representation into anchor and non-anchor, the attention map is checkerboard-like.}
  \label{fig:tecnick01_attn_map}
  \end{figure*}

To prove the effectiveness of our global context modules,
we provide more attention maps in Figure \ref{fig:kodim15_attn_map},
\ref{fig:kodim20_attn_map}, \ref{fig:tecnick01_attn_map}, \ref{fig:tecnick02_attn_map}.
\section{Influence of Resolution}
\begin{figure*}
  \centering
  \includegraphics[width=0.75\linewidth]
  {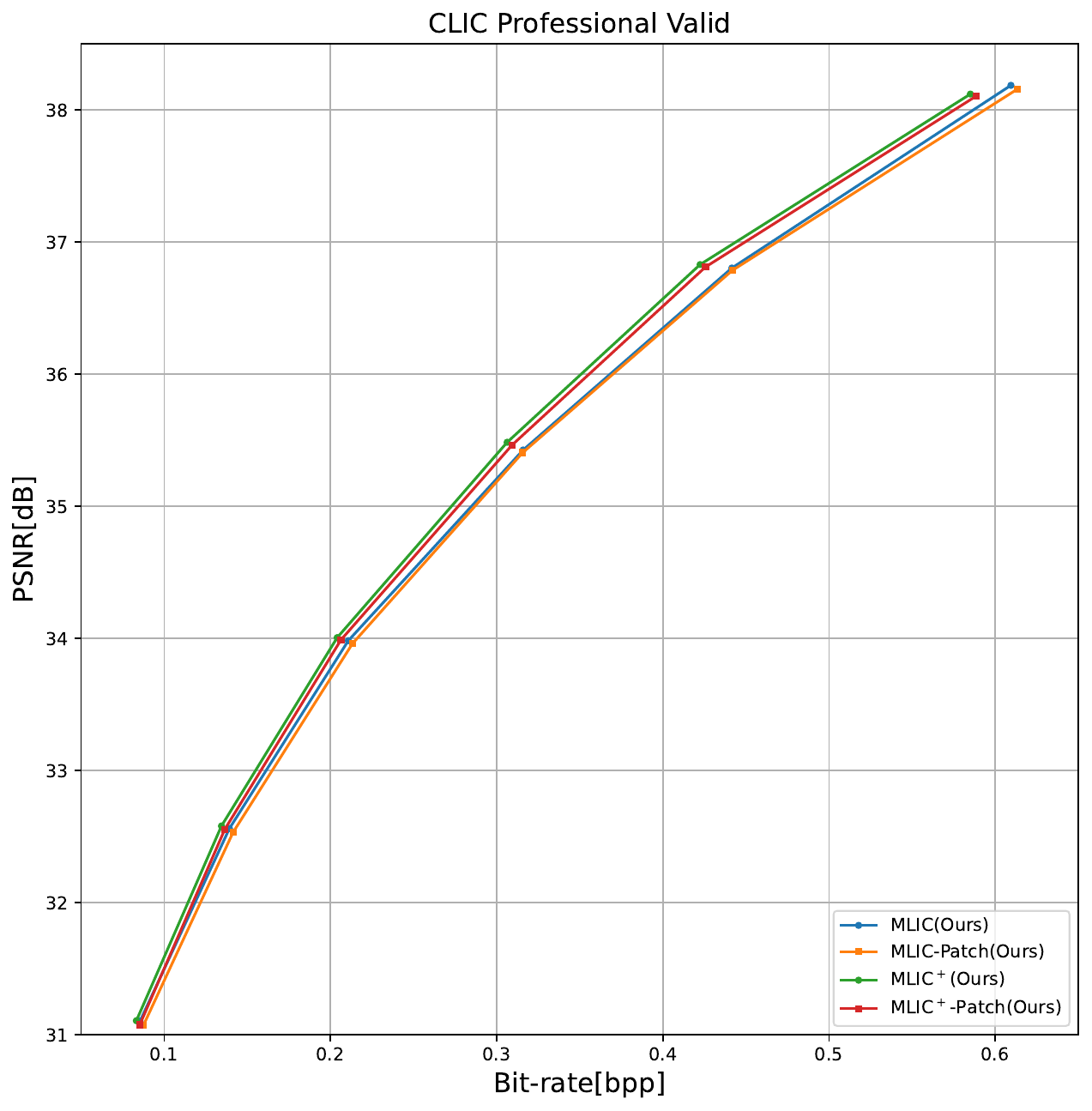}
  \caption{Rate-distortion performance comparison between compressing an image
  and compressing non-overlapped patches.
  We evaluate performance on CLIC-Professional Valid dataset.
  The size of each patch is $448\times 448$.}
  \label{fig:clicp_patch}
  \end{figure*}

\begin{figure*}
  \centering
  \includegraphics[width=0.85\linewidth]
  {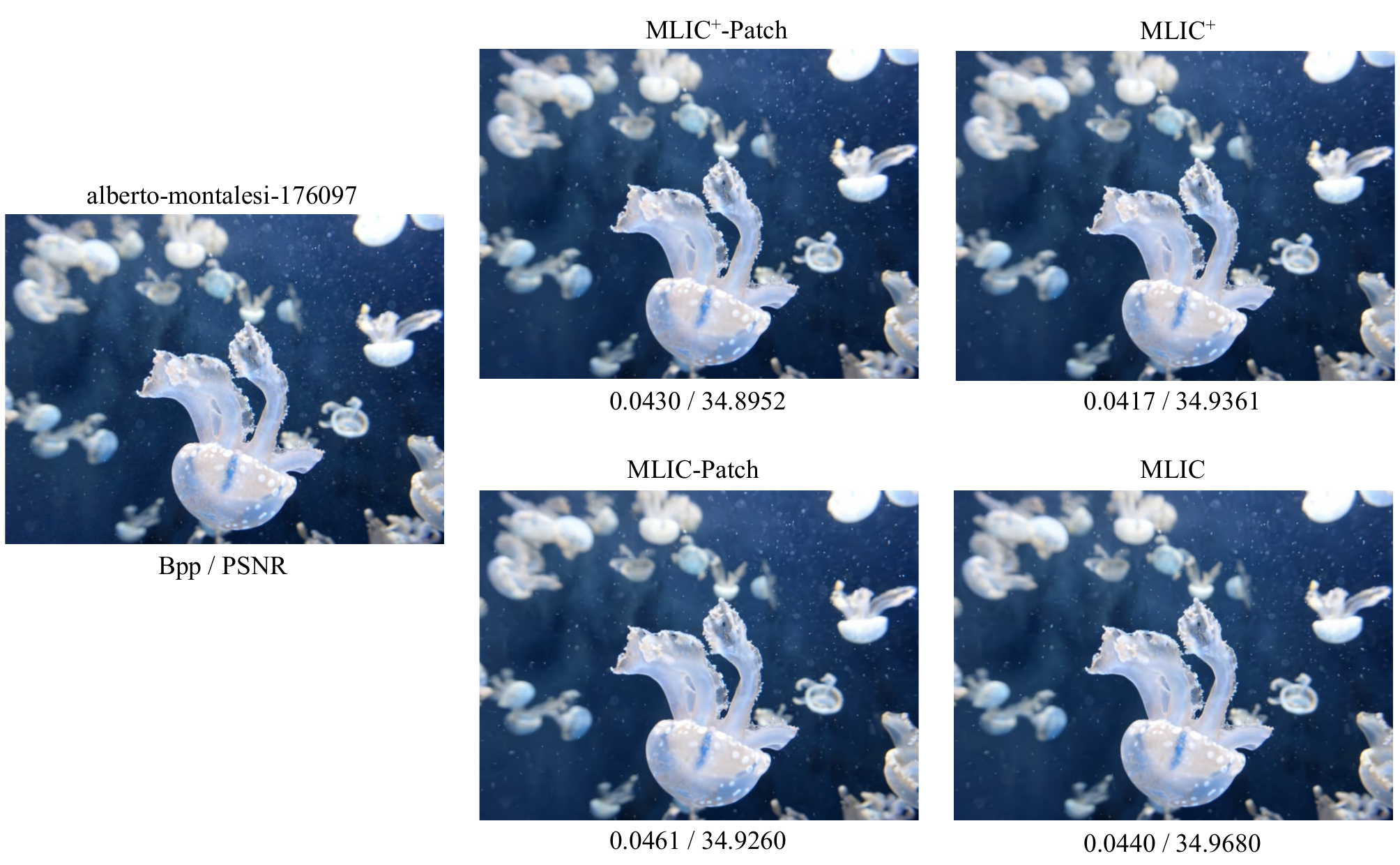}
  \caption{Visualization of the reconstructed image alberto-montalesi-176097.
  The metrics are [bpp↓/PNSR↑].}
  \label{fig:patch_visual}
  \end{figure*}

We are interested in the influence of resolution of images.
Our intra-slice global context module $g_{gc, intra}$ and
inter-slice global context module $g_{gc, inter}$ are based on
Transformers \cite{DBLP:conf/nips/VaswaniSPUJGKP17}.
Transformers are not translation invariant,
which means there is an inconsistency between training and testing.
Another reason is the computational complexity when compressing images with high resolution,
which means that we need to crop images with high resolution into patches.
We compare our MLIC and MLIC$^+$ on CLIC Professional Valid dataset \cite{clic2020dataset}.
We crop images into non-overlapped $448\times 448$ patches.
To avoid the influence of padding zeros, all images are cropped to multiples of $448$.
The results in illustrated in Figure \ref{fig:clicp_patch}.
When compressing non-overlapped patches, there is almost no performance degradation.
Maybe global correlations within $448\times 448$ patch are enough for conditional
probability estimation. Figure \ref{fig:patch_visual}
shows the example of compressing an image and compressing patches.
\section{Progressive Decoding Analysis}
\begin{figure*} [t!]
    \centering
    \subfloat{
     \includegraphics[scale=0.63]{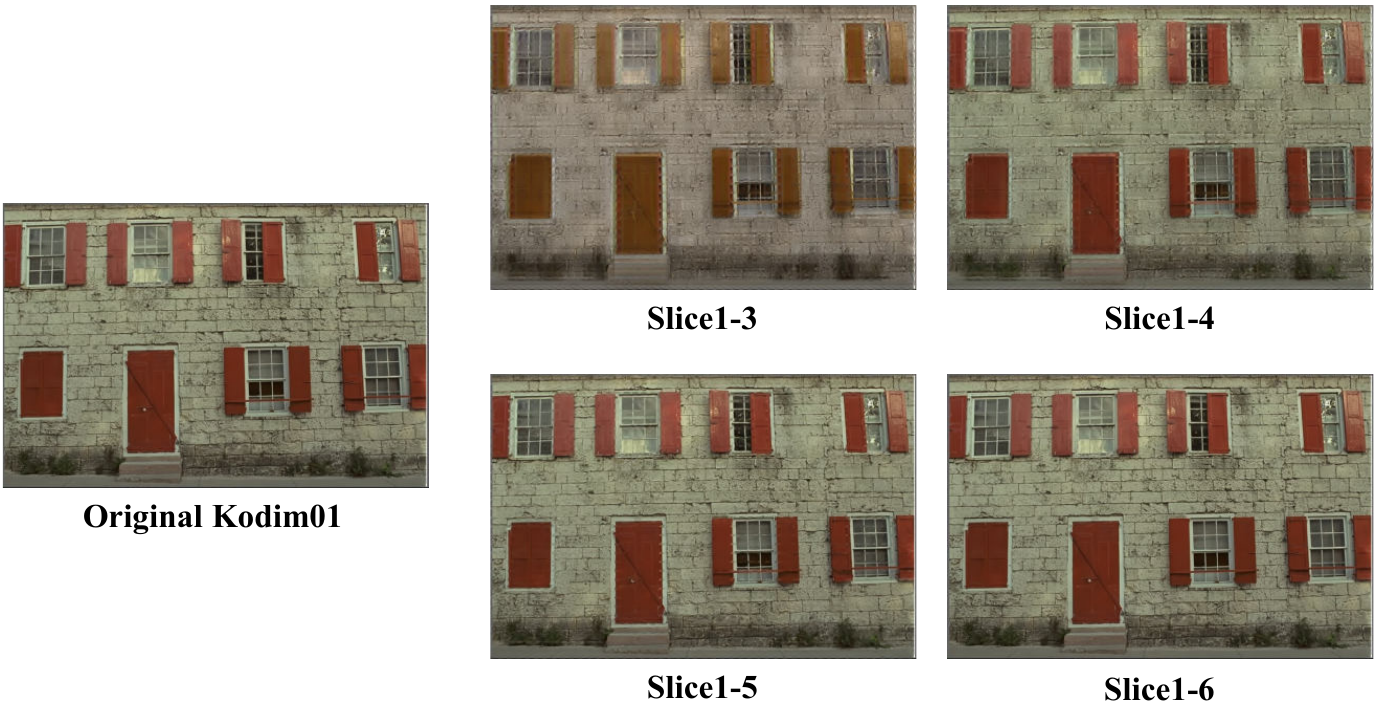}}\\
    \subfloat{
     \includegraphics[scale=0.63]{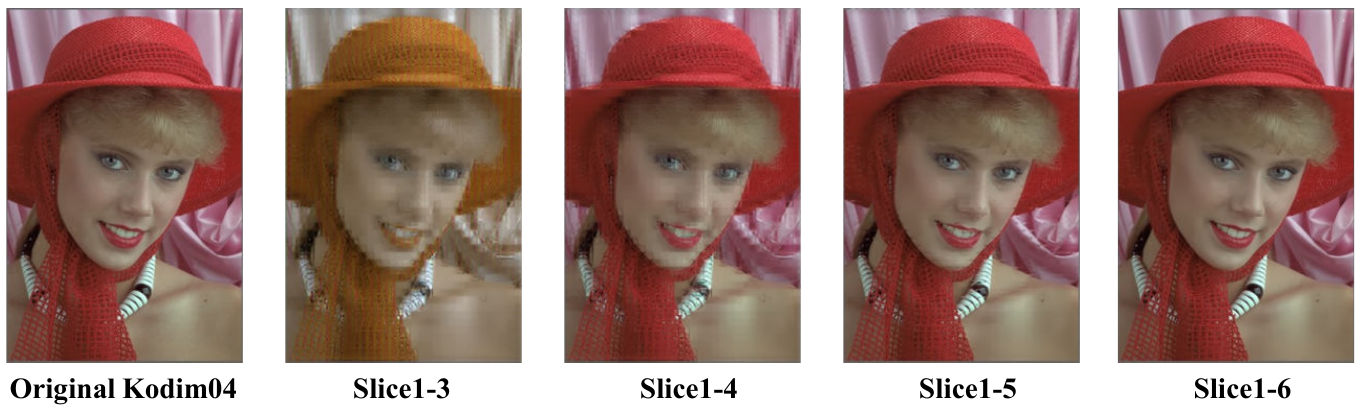}}\\
    \subfloat{
     \includegraphics[scale=0.63]{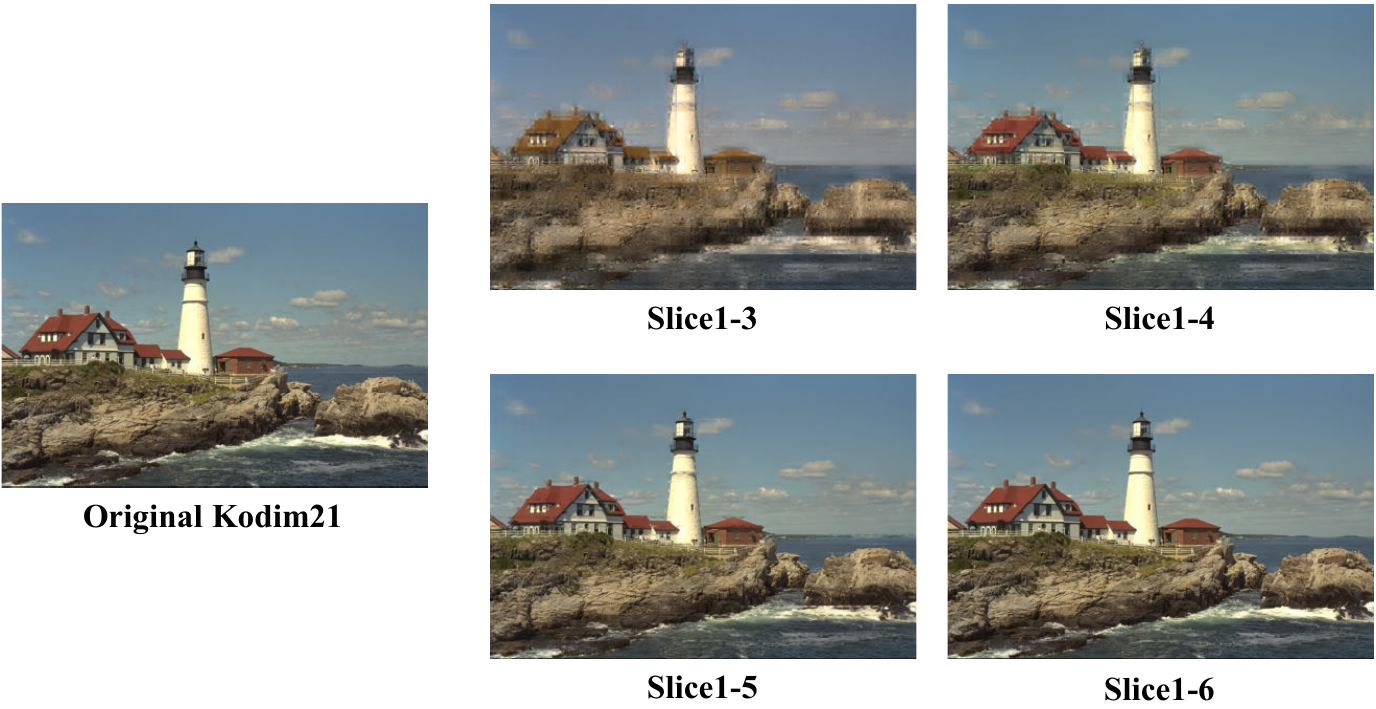}}
    \caption{Progressive Decoding Results of MLIC on Kodak dataset.
    We pad latent representation with zeros.}
    \label{fig:cgs}
   \end{figure*}

Learned Image Compression Models with channel-wise context model usually
support progressive decoding \cite{DBLP:conf/icip/MinnenS20, He_2022_CVPR}.
We also illustrate the progressive decoding results of our MLIC in Figure \ref{fig:cgs}.
However, the progressive decoding performance of MLIC is much worse than the
performance of separated optimized models. What's more, we find the performance
of progressive decoding is not stable. Sometimes progressive decoding leads to
unpleasant artifacts and noise or crashes. Progressive decoding results of our MLIC$^+$
is much worse. The problems can be attributed to our global spatial context modules.
It seems that the inter-slice global context model changes the information distribution
between slices. Compared with channel-wise context model, our global context modules
are extra constraints.\par
We think channel-wise context model is a good method for progressive decoding.
But when using channel-wise context model, we need to design a suitable method
to change the information between slices to optimize the performance \cite{ma2022deepfgs}.
   \begin{figure*}
   \centering
   \includegraphics[width=\linewidth]
   {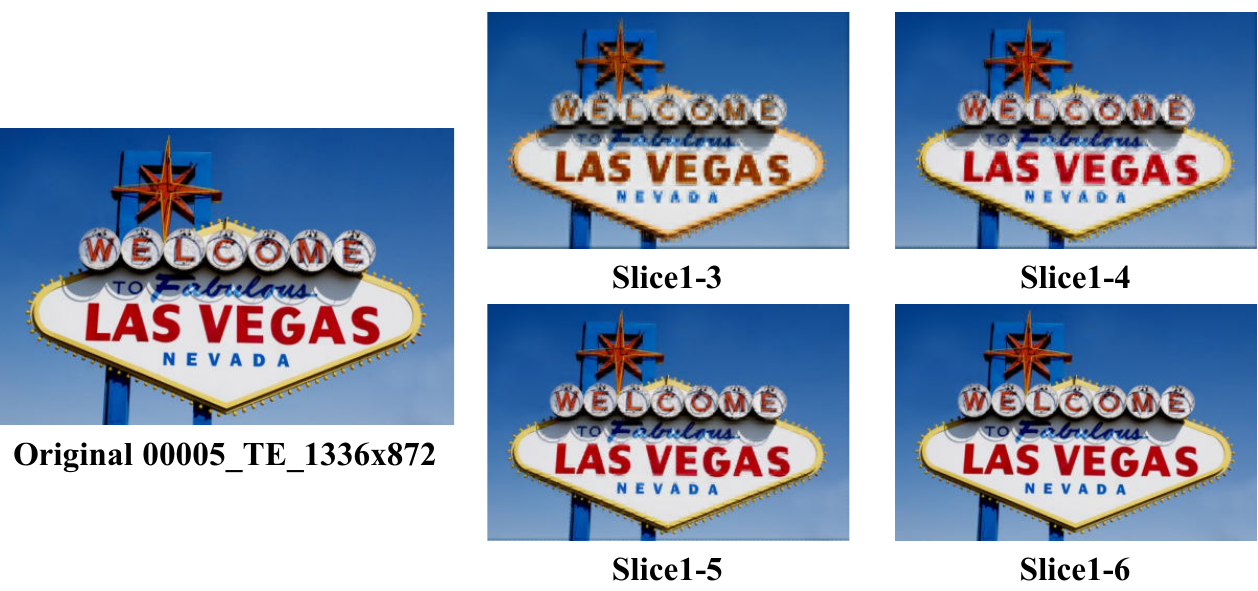}
   \caption{Progressive Decoding Results of MLIC on JPEGAI Test dataset.
   We pad latent representation with zeros.}
   \label{fig:jpegai_cgs}
   \end{figure*}
   \begin{figure*}
   \centering
   \includegraphics[width=\linewidth]
   {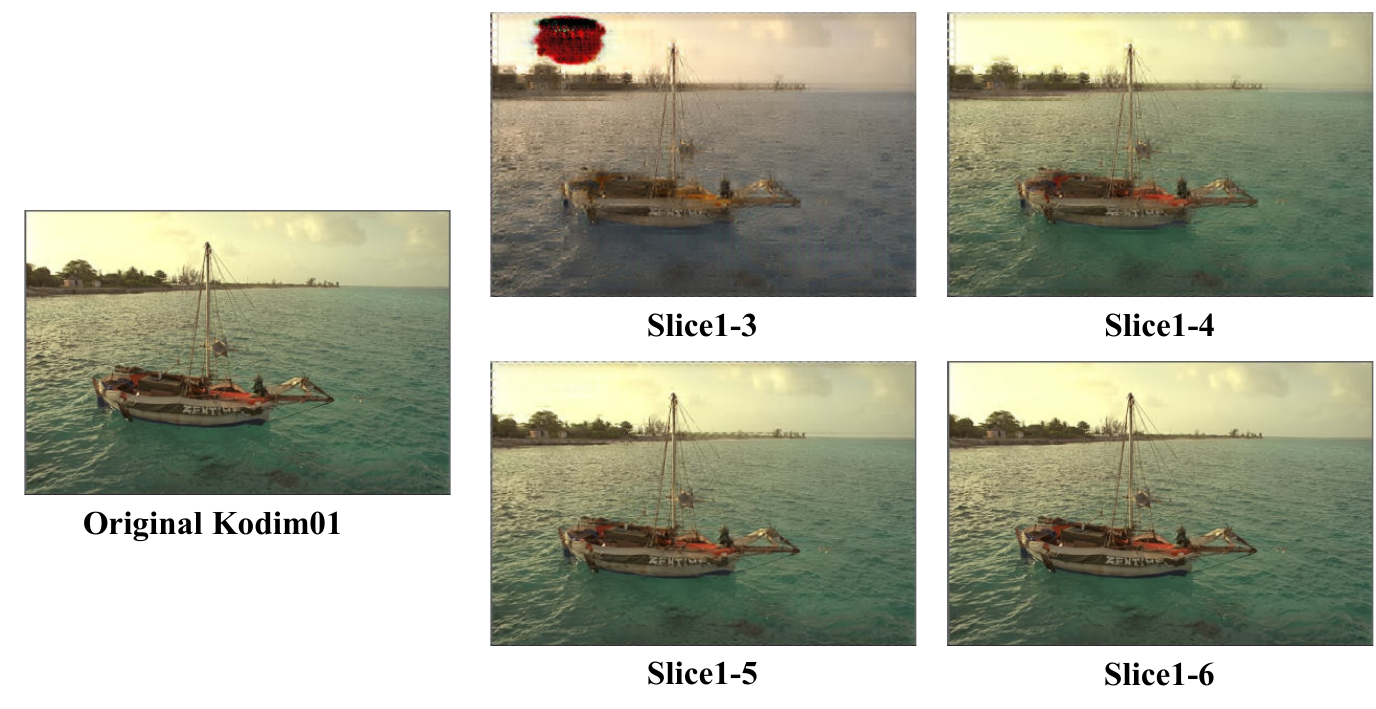}
   \caption{Examples of crashed progressive decoding.
   We pad latent representation with zeros.}
   \label{fig:crash_cgs}
   \end{figure*}
\bibliographystyle{ACM-Reference-Format}
\balance
\bibliography{mm23}
\end{document}